\documentclass[aps,prb,twocolumn,superscriptaddress,nofootinbib,floatfix,longbibliography]{revtex4-2}

\usepackage{amsmath,amssymb,bm}
\usepackage{graphicx}
\usepackage[colorlinks=true,citecolor=blue,linkcolor=blue,urlcolor=blue]{hyperref}
\usepackage[percent]{overpic}
\usepackage[export]{adjustbox}

\newcommand{\ket}[1]{|#1\rangle}

\newcommand{\ii}{\mathrm{i}}
\newcommand{\eps}{\varepsilon}
\newcommand{\Tstep}{T_{s}}
\newcommand{\UF}{U_F}

\newcommand{\QuICS}{
Joint Center for Quantum Information and Computer Science, NIST/University of Maryland, College Park, Maryland 20742, USA}
\newcommand{\JQI}{
Joint Quantum Institute, NIST/University of Maryland, College Park, Maryland 20742, USA}

\begin{document}

\title{Anomalous Boundary Modes in a Floquet Hyperbolic System}

\author{Ali Fahimniya}
\email{fahim@umd.edu}
\affiliation{\QuICS}
\affiliation{\JQI}

\author{Hossein Dehghani}
\altaffiliation{Current address: QuEra Computing Inc., Boston, MA 02135, USA}
\affiliation{\QuICS}
\affiliation{\JQI}
\author{Alicia J.  Koll\'ar}
\affiliation{\JQI}
\affiliation{Department of Physics, University of Maryland, College Park, MD 20742, USA}
\affiliation{Maryland Quantum Materials Center, Department of Physics, University of Maryland, College Park, MD 20742, USA}

\author{Alexey V. Gorshkov}
\affiliation{\QuICS}
\affiliation{\JQI}

\date{\today}

\begin{abstract}
We construct an anomalous Floquet topological phase on a negatively curved hyperbolic lattice. The model is a tight-binding Hamiltonian with a periodically repeated four-color edge-hopping sequence and a sublattice-staggered onsite potential step. The topological regime is reached near the limit in which a single hopping step transfers amplitude completely across an active edge, while the trivial regime is reached near the point where two full hops occur along an active edge during a single hopping step, returning the amplitude to its starting site. %
In finite open patches, the topological regime is characterized by bulk quasienergy gaps at $0$ and $\pi$ that are populated by in-gap states, in contrast to a trivial regime where these gaps remain empty. Using compact periodic lattices, we map the bulk $0$ and $\pi$ quasienergy gaps and identify the gapped regions connected to the trivial and anomalous open-boundary spectra. We diagnose the in-gap states as chiral boundary modes by their real-space dynamics. Finally, we introduce a small-boundary spectral-flow diagnostic based on punctured periodic hyperbolic lattices, which avoids the ambiguity associated with the extensive outer boundary of finite hyperbolic patches. This puncture-based diagnostic should be useful for studying other topological hyperbolic systems.

\end{abstract}

\maketitle

\section{Introduction}

Floquet engineering provides a route to designing band structures, gauge fields, and topological phases through periodic time dependence rather than through static material parameters~\cite{goldman_periodically_2014,bukov_universal_2015,eckardt_colloquium_2017,rudner_band_2020}. Early proposals showed that irradiation can induce Hall or topological-insulator responses in otherwise ordinary systems, including graphene and semiconductor quantum wells~\cite{oka_photovoltaic_2009,lindner_floquet_2011}. More generally, a periodically driven system with driving period $T$ is governed stroboscopically by a one-period unitary evolution operator, whose eigenphases define quasienergies modulo $2\pi/T$. This quasienergy periodicity gives Floquet band topology a structure that is richer than that of static Bloch bands~\cite{kitagawa_topological_2010,rudner_anomalous_2013,titum_anomalous_2016,nathan_topological_2015,roy_periodic_2017}. In particular, the anomalous Floquet insulator introduced by Rudner \textit{et al.} supports chiral edge modes even when the Chern numbers of all Floquet bands vanish~\cite{rudner_anomalous_2013}. Such anomalous edge transport is not only a theoretical possibility: photonic Floquet topological insulators and anomalous Floquet edge modes have been observed in periodically modulated waveguide lattices~\cite{rechtsman_photonic_2013,maczewsky_observation_2017,mukherjee_experimental_2017}. Related anomalous-Floquet physics has also been realized in acoustic, ultracold-atom, and nanophotonic resonator platforms~\cite{peng_experimental_2016,wintersperger_realization_2020,afzal_realization_2020}. Periodic driving therefore separates the topology of the full time evolution from the topology of individual static-like bands, making it a natural setting for engineering boundary motion directly.

Hyperbolic lattices provide a complementary setting in which boundary physics is unusually prominent. A regular tiling may be denoted by a Schl\"afli symbol $\{p,q\}$, where $p$ is the number of sides of each polygonal face and $q$ is the number of polygons meeting at each vertex~\cite{a.blatov_vertex_2010}. Hyperbolic tilings obey $(p-2)(q-2)>4$, in contrast to Euclidean regular tilings where equality holds. A key consequence of negative curvature is that finite patches of hyperbolic lattices have an extensive boundary: the number of boundary sites remains a finite fraction of the total number of sites even as the system grows. This feature is especially relevant for topological systems, whose most visible physical signatures are often boundary modes. It also complicates standard diagnostics because the usual separation between a large bulk and a negligible boundary is absent in open hyperbolic flakes.

The experimental motivation for hyperbolic tight-binding models has grown rapidly. Networks of superconducting coplanar waveguide resonators have realized effective hyperbolic lattices for microwave photons~\cite{kollar_hyperbolic_2019}. Classical electric-circuit platforms have independently emulated hyperbolic space on a circuit board and have been used to engineer hyperbolic matter with tunable complex phases~\cite{lenggenhager_simulating_2022,chen_hyperbolic_2023}. Related circuit experiments have observed boundary-dominated topological states and higher-order modes in engineered hyperbolic lattices~\cite{zhang_observation_2022}. Photonic platforms are also emerging: coupled optical ring resonators on silicon chips have realized hyperbolic photonic topological insulators, and programmable coupled-resonator photonics has been proposed as a scalable route for emulating hyperbolic-lattice wave dynamics~\cite{huang_hyperbolic_2024a,park_scalable_2024}. These developments make hyperbolic lattices a realistic synthetic-matter platform rather than only a mathematical generalization of Euclidean crystalline systems.

On the theory side, hyperbolic band theory differs sharply from ordinary Bloch theory because the relevant translation groups are nonabelian. Work on hyperbolic Bloch theory, automorphic Bloch theorems, hyperbolic crystallography, and higher-dimensional representations has developed tools for describing spectra of infinite or compactified hyperbolic lattices~\cite{maciejko_hyperbolic_2021,maciejko_automorphic_2022,boettcher_crystallography_2022,cheng_band_2022}. Complementary progress on converging periodic boundary conditions and supercell constructions has provided practical routes to approximating thermodynamic-limit spectra and non-Abelian Bloch sectors of hyperbolic lattices~\cite{lux_converging_2023,lenggenhager_nonabelian_2023}. Topological hyperbolic systems have also been explored in several forms, including hyperbolic analogues of quantum spin Hall systems~\cite{yu_topological_2020}, Chern insulators on the $\{8,3\}$ lattice~\cite{liu_chern_2022}, hyperbolic Haldane and Kane--Mele models~\cite{urwyler_hyperbolic_2022}, higher-order topological hyperbolic phases~\cite{liu_higherorder_2023,tao_higherorder_2023,zhang_hyperbolic_2023}, linear Chern responses~\cite{sun_topological_2024}, and nonreciprocal hyperbolic scattering networks with anomalous and Chern chiral edge modes~\cite{chen_anomalous_2024}. This body of work establishes that hyperbolic geometry is not a passive background: the extensive boundary, noncommutative translation structure, and available synthetic implementations all reshape the usual questions of band topology and bulk-boundary correspondence.

Here we construct a Hamiltonian Floquet model on the $\{8,3\}$ hyperbolic lattice, in the spirit of the Rudner hopping protocol~\cite{rudner_anomalous_2013} (see Fig.~\ref{fig:model}). The lattice consists of regular octagons with three octagons meeting at each vertex. We use a periodic 16-site coloring of the lattice edges into four classes (four colors). During the drive, hopping is activated on these four edge classes in a repeated sequence, followed by a sublattice-staggered onsite potential. We identify a topological regime near the perfect-hopping limit, where one active step transfers amplitude completely across an edge, and a trivial regime near the point where two full hops occur along an active edge during a single hopping step, returning the amplitude to its starting site. %
We diagnose the resulting anomalous Floquet regime through the density of states of finite flakes and periodic lattices, chiral boundary wave-packet dynamics, and finally, spectral flow in punctured compact hyperbolic geometries.

The remainder of this paper is organized as follows. Section~\ref{sec:model} introduces the hyperbolic $\{8,3\}$ lattice, its four-color edge decomposition, and the corresponding Floquet drive, including the real-space dynamics at the perfect-hopping point. Section~\ref{sec:results} presents the numerical characterization of the resulting phases. We first compare the quasienergy densities of states of finite open patches in Sec.~\ref{subsec:dos}, and then map the bulk $0$ and $\pi$ gaps using periodic lattices in Sec.~\ref{subsec:phase_diagram}. Section~\ref{subsec:boundary_dynamics} demonstrates the chiral propagation of a boundary wave packet, while Sec.~\ref{subsec:spectral_flow} introduces a controlled internal boundary in a compact periodic lattice and diagnoses the anomalous Floquet phase through puncture-boundary spectral flow. Finally, Sec.~\ref{sec:discussion} summarizes the implications of these results and discusses generalizations to other hyperbolic lattices, real-space topological diagnostics, and possible experimental realizations.

\section{Model}
\label{sec:model}

\subsection{Hyperbolic lattice and periodic edge coloring}
\label{subsec:lattice_coloring}

\begin{figure}[t]
    \centering
    \includegraphics[width=1\columnwidth]{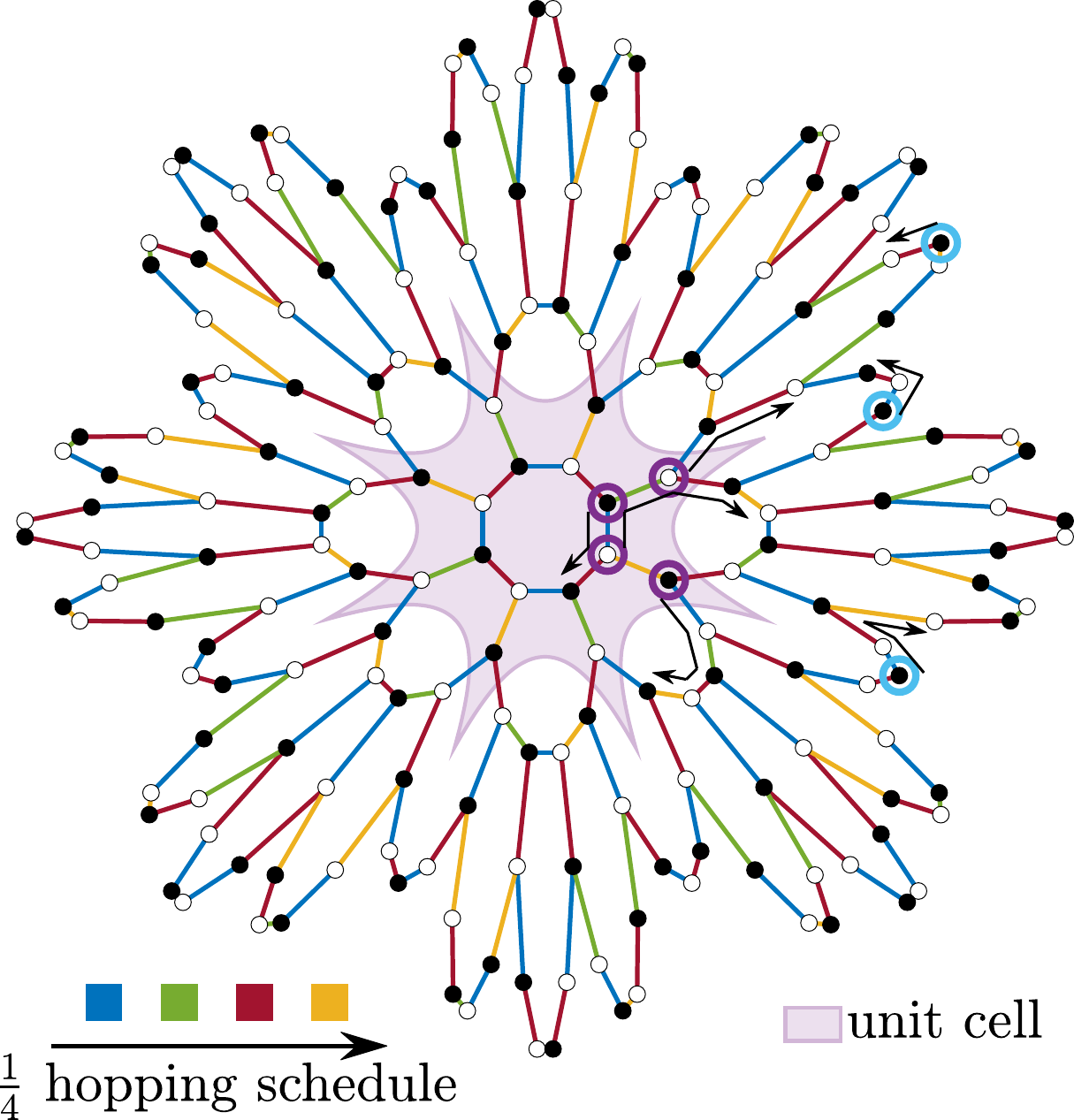}

    \caption{
    \textit{Hyperbolic $\{8,3\}$ lattice and four-color hopping protocol.} The lattice consists of regular octagons with three octagons meeting at each bulk vertex. Black and white vertices denote the two sublattices. The shaded region marks a 16-site fundamental domain of the colored lattice. The lattice edges are divided into four color classes, blue, green, red, and orange. The four-step color cycle blue–green–red–orange is repeated four times and is followed by a sublattice-staggered onsite potential step. Purple-circled vertices mark the four inequivalent bulk vertex types of the colored lattice, while cyan-circled vertices show representative boundary-propagating sites. Arrows show perfect-hopping point motion of each site over one quarter of the hopping schedule.
    }
    \label{fig:model}
\end{figure}

We consider the regular hyperbolic tessellation $\{8,3\}$, whose faces are octagons and whose bulk vertices have coordination number three. Finite open-boundary systems are generated by starting from a central octagon and adding shells of neighboring octagons. The first shell is the central octagon; the next shell consists of octagons adjacent to it; subsequent shells are defined recursively by adding octagons adjacent to the previous shells but not already included. Boundary sites of an open patch are defined graph-theoretically as vertices whose degree is reduced from the bulk value three to degree two. This definition distinguishes sites affected by the finite boundary truncation, which are candidates for supporting the boundary states discussed below, from degree-three sites that primarily contribute to bulk states.

The hopping protocol is based on a periodic coloring of the edges. The colored lattice has a 16-site fundamental domain, shown schematically in Fig.~\ref{fig:model}. This unit cell is associated with the hyperbolic crystallographic description of the $\{8,3\}$ lattice~\cite{boettcher_crystallography_2022}. Every edge belongs to exactly one of four color classes, denoted
\begin{equation}
E_b,\quad E_g,\quad E_r,\quad E_o,
\end{equation}
corresponding to blue, green, red, and orange edges. The blue and red edges each form a perfect matching, i.e., each vertex is at the end of exactly one blue and exactly one red edge. The green and orange edges together form a third perfect matching. %
Consequently, each bulk vertex is incident on one blue edge, one red edge, and one edge that is either green or orange. The green and orange colors alternate spatially: after traversing a green edge and then a blue or red edge, one arrives at a vertex whose green-or-orange matching edge is orange, and conversely with green and orange interchanged. The coloring, along with the sublattice-staggered onsite potential defined below, breaks the full vertex equivalence of the uncolored $\{8,3\}$ lattice. As indicated by the purple-circled vertices in Fig.~\ref{fig:model}, the 16 sites in the fundamental domain reduce to four inequivalent vertex types under the residual fourfold rotation symmetry.

The graph is bipartite, and we denote the two sublattices by $A$ and $B$, indicated by white and black vertices in Fig.~\ref{fig:model}, respectively. The bipartite structure is used in the onsite potential step of the drive below.

\subsection{Floquet hopping protocol}
\label{subsec:floquet_protocol}

For each color class $\mu=b,g,r,o$, we define a nearest-neighbor hopping Hamiltonian
\begin{equation}
H_\mu
=
-J\sum_{\langle ij\rangle\in E_\mu}
\left(c_i^\dagger c_j+c_j^\dagger c_i\right),
\qquad J>0,
\label{eq:Hmu}
\end{equation}
where $c_i^\dagger$ creates a particle on site $i$, and we set $\hbar=1$. The hopping is real and thus reciprocal in the main model. Peierls phases are introduced only in the spectral-flow calculations discussed in Sec.~\ref{subsec:spectral_flow}.

The hopping part of the drive consists of a 16-step sequence obtained by repeating the four-step blue--green--red--orange hopping schedule four times.
Each step has duration $\Tstep$. The hopping sequence is followed by a sublattice-staggered onsite potential step
\begin{equation}
H_\delta
=
\delta \sum_i \eta_i c_i^\dagger c_i,
\qquad
\eta_i=
\begin{cases}
+1, & i\in A,\\
-1, & i\in B.
\end{cases}
\label{eq:HDelta}
\end{equation}
The full Floquet period is therefore
\begin{equation}
T=17\Tstep.
\end{equation}

Let
\begin{equation}
U_\mu=e^{-\ii H_\mu\Tstep},
\qquad
U_\delta=e^{-\ii H_\delta\Tstep}.
\end{equation}
The unitary evolution during a Floquet period is determined by the Floquet operator that is
\begin{equation}
\UF
=
U_\delta
\left(U_o U_r U_g U_b\right)^4.
\label{eq:UF}
\end{equation}
Because the Hamiltonian is periodic in time, $H(t+T)=H(t)$, the eigenvectors of $\UF$ are Floquet states and the eigenphases of $\UF$ define quasienergies,
\begin{equation}
\UF \ket{\psi_\alpha}
= e^{-\ii \eps_\alpha T}\ket{\psi_\alpha},
\qquad
\eps_\alpha T\in(-\pi,\pi].
\label{eq:quasienergy}
\end{equation}
The quasienergy is defined only modulo $2\pi/T$, so the spectrum is naturally defined on a circle rather than on a line. As a result, gaps at $\eps T=0$ and $\eps T=\pi$ are both meaningful. In an anomalous Floquet phase, chiral boundary states can traverse these quasienergy gaps even when static-like band invariants do not distinguish the bands~\cite{rudner_anomalous_2013}. We therefore focus on the quasienergy spectrum of $\UF$ and on the spatial structure and dynamics of states inside the $0$ and $\pi$ gaps.

The perfect-hopping value of a single active bond is
\begin{equation}
J\Tstep=\frac{\pi}{2},
\label{eq:perfect_point}
\end{equation}
for which the two-site hopping unitary transfers amplitude completely across the active bond, up to a phase. We identify the topological regime near this value. The trivial regime is obtained near the no-transfer points, including $J\Tstep=0$ and, by the single-bond periodicity of the hopping pulse, $J\Tstep=\pi$.

At the perfect-hopping point, which we also refer to as the perfect-point, the hopping part of the evolution reduces to a deterministic map of site amplitudes. The local motion can be read off from the four inequivalent bulk sites highlighted by purple circles in Fig.~\ref{fig:model}. During one quarter of the hopping schedule, i.e., one blue--green--red--orange sequence, two of these inequivalent sites undergo two hops, while the other two undergo four hops. In all four cases, the hops circulate clockwise around bulk octagons. After the full 16-step hopping protocol, the first pair of bulk-site types completes one loop around an octagon, while the second pair completes two loops. The onsite potential step only adds sublattice-dependent phases and does not change this site map.

The boundary behaves differently. Some degree-two boundary sites still complete local loops around octagons if the octagon selected by the hopping schedule remains intact after the finite boundary truncation. However, when the octagon that would close the local loop has been removed by the boundary cut, the same perfect-point dynamics translates the amplitude along the outer boundary instead. The cyan-circled sites and black arrows in Fig.~\ref{fig:model} show representative examples over one quarter of the hopping schedule. These propagating boundary sites move counterclockwise, opposite to the clockwise circulation of the bulk loops. We use this perfect-point motion to distinguish degree-two boundary sites that participate in propagating boundary motion from degree-two sites that close local loops. For the shell-by-shell construction of finite patches considered here, we estimate that approximately $43\%$ of all sites participate in propagating boundary motion rather than local perfect-point loops in the thermodynamic limit. This estimate is obtained by comparing the corresponding fractions for patches containing four, five, and six shells and extrapolating their finite-size trend.

\section{Results}
\label{sec:results}

\subsection{Quasienergy density of states}
\label{subsec:dos}

\begin{figure}[t]
    \centering

    \includegraphics[width=1\columnwidth]{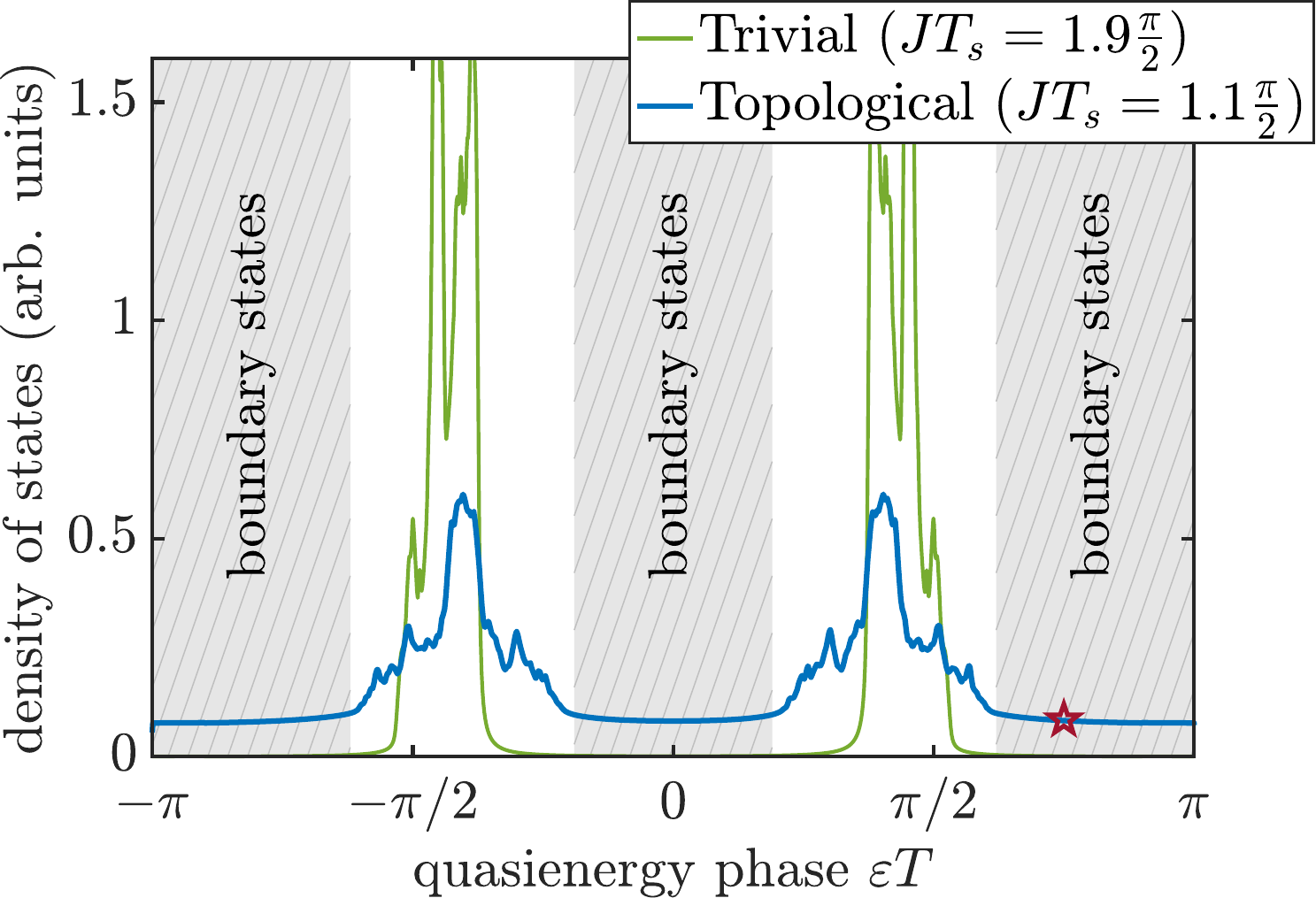}

    \caption{
    \textit{Quasienergy density of states of a six-shell open patch with $N=10800$ sites.} The quasienergy phase $\eps T$ is defined by $U_F|\psi_\alpha\rangle=e^{-i\eps_\alpha T}|\psi_\alpha\rangle$. The green curve, $J\Tstep=1.9\pi/2$, shows the trivial regime with gaps at $\eps T=0$ and $\pi$. The blue curve, $J\Tstep=1.1\pi/2$, shows the topological regime, where both quasienergy gaps, shaded in gray, are populated by boundary states. Both curves use ${\delta\Tstep=0.8\pi/2}$. The red star indicates the quasienergy of the in-gap state used for boundary dynamics in Fig.~\ref{fig:boundary_dynamics}.
    }
    \label{fig:dos}
\end{figure}

We first compare representative open-boundary spectra in the topological and trivial regimes. Figure~\ref{fig:dos} shows the quasienergy density of states for a six-shell open patch containing $N=10800$ sites. The boundary, defined by reduced graph degree, contains $N_\partial=6240$ sites, so that approximately $58\%$ of the sites lie on the open boundary. This extensive boundary fraction is a characteristic feature of finite hyperbolic patches and motivates the complementary diagnostics below.

The density of states is obtained by exact diagonalization of the finite-system Floquet operator in Eq.~\eqref{eq:UF}. To produce the continuous curves shown in Fig.~\ref{fig:dos}, each quasienergy is broadened by a Lorentzian with width $\eta=8\times10^{-3}$ in units of $\eps T$. The green curve corresponds to
\begin{equation}
J\Tstep=1.9\frac{\pi}{2},
\end{equation}
which is close to the trivial no-transfer point $J\Tstep=\pi$. In this regime the spectrum forms two quasienergy bands separated by clear gaps at $\eps T=0$ and $\eps T=\pi$. The gap at $0$ is opened by the finite sublattice-staggered onsite potential in Eq.~\eqref{eq:HDelta}. For both curves in Fig.~\ref{fig:dos}, the onsite pulse strength is ${\delta\Tstep=0.8\pi/2}$.

The blue curve corresponds to
\begin{equation}
J\Tstep=1.1\frac{\pi}{2},
\end{equation}
which is close to the perfect-hopping point. In this regime, the same bulk gaps are populated by in-gap states. The density of states alone does not establish the spatial character of these states. We identify them as boundary-localized through the wave-packet dynamics in Sec.~\ref{subsec:boundary_dynamics} and through the spectral-flow diagnosis discussed in Sec.~\ref{subsec:spectral_flow}.

The open-boundary density of states therefore identifies the two representative regimes used in the rest of the paper. To determine where these regimes sit relative to bulk gap closings, we next compute the $0$ and $\pi$ gaps on a compact periodic $\{8,3\}$ lattice, where the spectrum is not contaminated by the extensive boundary of an open hyperbolic patch.

\subsection{Bulk gap phase diagram}
\label{subsec:phase_diagram}

\begin{figure}[t]
    \centering
    \includegraphics[width=\columnwidth]{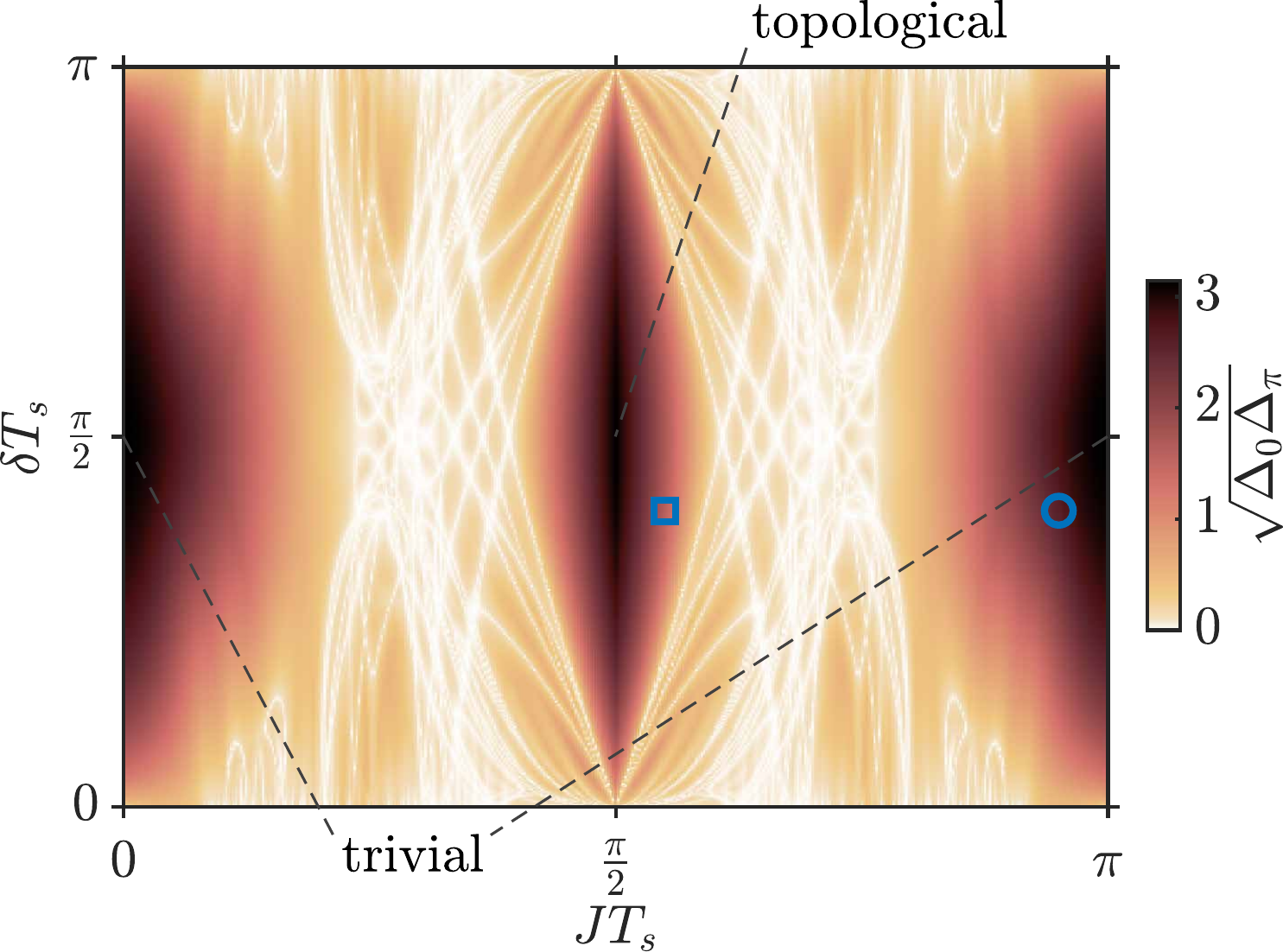}
    
    \caption{
    \textit{Bulk quasienergy gap phase diagram on a $2048$-site periodic $\{8,3\}$ lattice.} The color scale shows the geometric mean $\bar{\Delta}=\sqrt{\Delta_0\Delta_\pi}$, where $\Delta_0$ and $\Delta_\pi$ are the quasienergy-phase gaps across $\eps T=0$ and $\eps T=\pi$, respectively. Dark regions indicate that both bulk gaps are resolved in the finite periodic spectrum, while light regions indicate that at least one of the two gaps is numerically small. The blue square and circle correspond to the two open-boundary spectra shown in Fig.~\ref{fig:dos}: the blue square at $J\Tstep=1.1\pi/2$ lies in the anomalous Floquet region, while the blue circle at $J\Tstep=1.9\pi/2$ lies in the trivial region. Both marked points use $\delta \Tstep=0.8\pi/2$.
    }
    \label{fig:phase_diagram}
\end{figure}

The open-boundary density of states in Fig.~\ref{fig:dos} distinguishes the two representative parameter points used throughout this work. However, because a finite hyperbolic patch has an extensive boundary, the open-boundary spectrum alone does not cleanly locate the bulk gap closings that separate distinct regimes. To isolate the bulk spectrum, we compute the quasienergy gaps on a compact periodic $\{8,3\}$ lattice, following the strategy used in hyperbolic Hofstadter calculations to avoid open-boundary contamination of the bulk spectrum~\cite{stegmaier_universality_2022}. The periodic lattice used here has $N=2048$ sites and is taken from the list of trivalent symmetric graphs of Conder and Dobcs\'anyi~\cite{conder_trivalent_2002}\footnote{Reference~\cite{conder_trivalent_2002} includes methodology and graphs up to 768 vertices. Graphs with up to 10,000 vertices can be found \href{https://www.math.auckland.ac.nz/~conder/symmcubic10000list.txt}{here}.}. It realizes a closed finite quotient of the $\{8,3\}$ tiling on a genus-$129$ manifold and has no open boundary.

\begin{figure*}[t]
    \centering
    \includegraphics[width=0.95\textwidth]{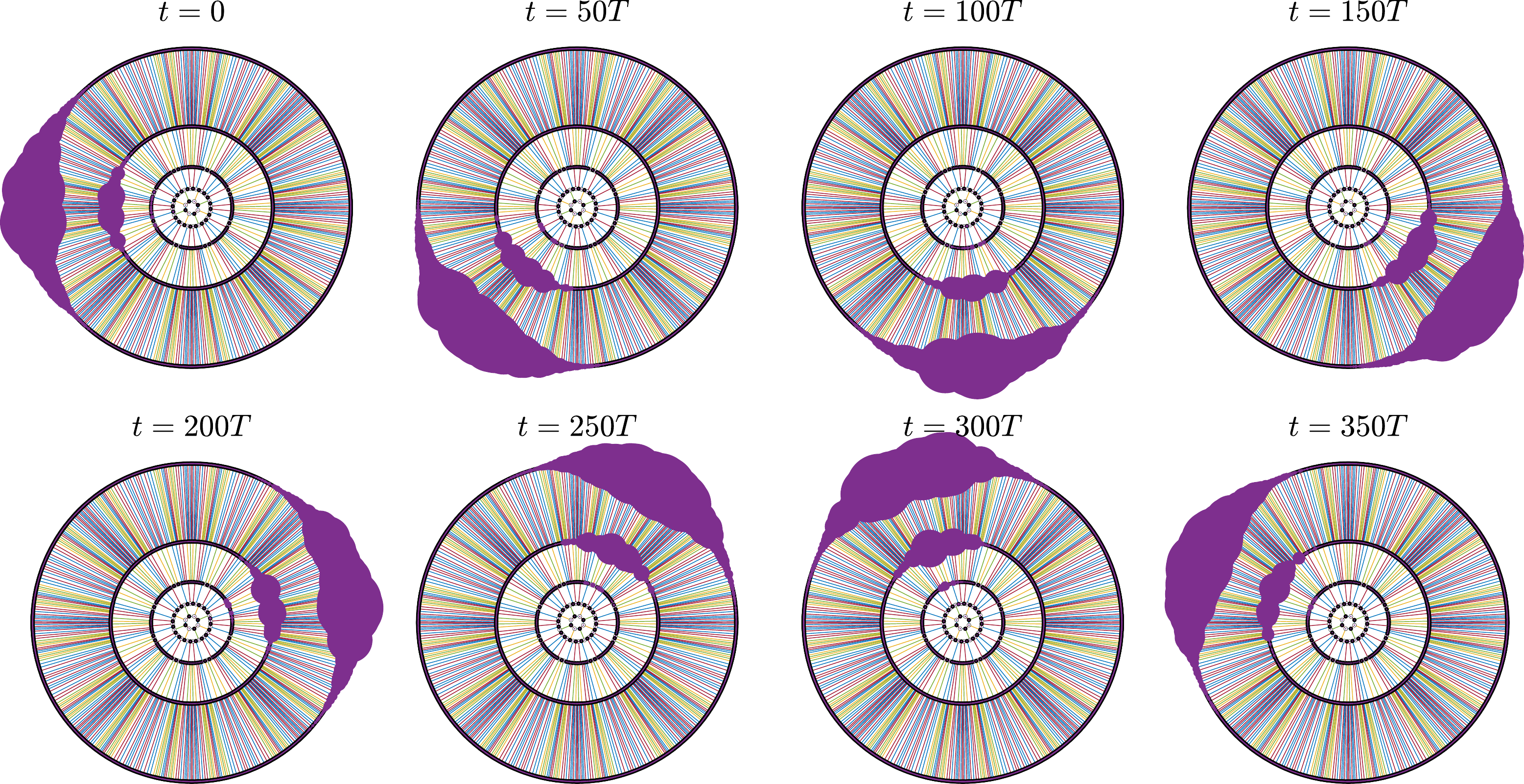}
    \caption{
    \textit{Stroboscopic evolution of a boundary wave packet in the anomalous Floquet regime, $J\Tstep=1.1\pi/2$.} The system is a five-shell open $\{8,3\}$ patch with $N=2888$ sites. This is a schematic radial plot where the sites on each shell are plotted on a circle. The initial state is obtained from the in-gap boundary state at the quasienergy marked by the red star in Fig.~\ref{fig:dos} by applying the angular envelope in Eq.~\eqref{eq:initial_packet}. The area of purple disks are proportional to $|\psi_i(t)|^2$ and are rescaled independently at each time so that the largest marker has a fixed display area; disk sizes should therefore not be compared quantitatively across panels. The panels show $t=0,50T,100T,150T,200T,250T,300T,350T$. The wave packet remains localized near the outer boundary and propagates counterclockwise around the sample.
    }
    \label{fig:boundary_dynamics}
\end{figure*}

For each pair of drive parameters $(J\Tstep,\delta \Tstep)$, we diagonalize the Floquet operator on this periodic lattice and sort the eigenphases $\theta_\alpha=\eps_\alpha T$ in the Floquet zone $(-\pi,\pi]$. We define $\Delta_0$ as the quasienergy-phase spacing across $\theta=0$, i.e., the difference between the smallest positive eigenphase and the largest negative eigenphase. The $\pi$ gap is defined across the Floquet-zone edge,
\begin{equation}
    \Delta_\pi = 2\pi+\theta_1-\theta_N ,
\end{equation}
where $\theta_1$ and $\theta_N$ are the smallest and largest sorted eigenphases. The quantity plotted in Fig.~\ref{fig:phase_diagram} is the geometric mean
\begin{equation}
    \bar{\Delta}=\sqrt{\Delta_0\Delta_\pi}.
\end{equation}
This quantity is nonzero only when both Floquet gaps are resolved in the finite periodic spectrum, and therefore identifies the parameter regions where the gap topology can be meaningfully compared.

Figure~\ref{fig:phase_diagram} shows extended regions in which both the $0$ and $\pi$ bulk gaps are resolved, separated by regions where at least one of the two gaps becomes numerically small. The individual gap maps, $\Delta_0$ and $\Delta_\pi$, are shown in Appendix~\ref{app:individual_gaps}. The two parameter points used in Fig.~\ref{fig:dos} lie in distinct gapped regions. The point near complete transfer across an active bond, $J\Tstep=1.1\pi/2$ with $\delta \Tstep=\pi/4$, lies in the anomalous Floquet region. In the corresponding open-boundary system, both quasienergy gaps contain in-gap states identified as boundary states by the dynamical and spectral-flow diagnostics in Secs.~\ref{subsec:boundary_dynamics} and \ref{subsec:spectral_flow}. The point near two transfers during one hopping pulse, $J\Tstep=1.9\pi/2$ with $\delta \Tstep=\pi/4$, lies in the trivial region, where the open-boundary spectrum retains empty gaps.

Within the gapped regions resolved in this finite periodic system, we find no numerical evidence for an additional Chern-band regime. In a Floquet system with two isolated quasienergy band groups, a nonzero Chern number for one band group would be reflected in a difference between the net chiral edge content of the two quasienergy gaps. Instead, the open-boundary spectra distinguish only the two regimes discussed above: a trivial regime with no boundary states in either gap, and an anomalous Floquet regime in which boundary states appear in both gaps.

\subsection{Chiral boundary dynamics}
\label{subsec:boundary_dynamics}

Having identified the bulk-gapped regions in Fig.~\ref{fig:phase_diagram}, we now return to an open finite patch and examine the real-space dynamics of a boundary wave packet in an open five-shell patch with $N=2888$ sites. The smaller system is used for visual clarity. Because no conventional translation-invariant momentum-space description is available for the boundary of this finite hyperbolic patch, we construct the localized wave packet directly in real space~\cite{boettcher_crystallography_2022,cheng_band_2022}. We choose an eigenstate associated with an in-gap boundary mode and form a localized wave packet by multiplying its site amplitudes by an angular Gaussian envelope,
\begin{equation}
\psi_i(0) \propto \phi_i \exp\left[-\frac{d_\theta(\theta_i,\theta_0)^2}{\sigma_\theta^2}\right],
\label{eq:initial_packet}
\end{equation}
where $\phi_i$ is the chosen boundary eigenstate, $\theta_i$ is the polar angle of site $i$ in the Poincar\'e disk representation, and
\begin{equation}
d_\theta(\theta,\theta_0)=\min_{n\in\mathbb{Z}}|\theta-\theta_0+2\pi n|
\end{equation}
is the periodic angular distance between $\theta$ and $\theta_0$ on the
circle. The center angle is denoted by $\theta_0$ and the width $\sigma_\theta$ is chosen to be $0.5$. The state is then evolved stroboscopically,
\begin{equation}
\ket{\psi_i(nT)}=\UF^n\ket{\psi_i(0)}.
\label{eq:stroboscopic_evolution}
\end{equation}

Figure~\ref{fig:boundary_dynamics} shows the resulting evolution at $J\Tstep=1.1\pi/2$, using the in-gap boundary eigenstate with a quasienergy marked by the red star in Fig.~\ref{fig:dos} to construct the initial wave packet. Other in-gap boundary eigenstates give qualitatively similar dynamics, differing mainly in their group velocity. The area of each purple disk is proportional to the site probability density $|\psi_i(nT)|^2$. Over the time interval shown, the wave packet remains confined to the outer boundary region, with no visible weight propagating into the inner shells. At the same time, the packet moves counterclockwise around the sample. This chiral boundary propagation provides a real-space dynamical signature of the in-gap boundary states inferred from the density of states. The unfiltered boundary state, together with a representative bulk-band state and the corresponding bulk wave-packet dynamics, is shown in Appendix~\ref{app:wavepacket_dynamics}.

\subsection{Flux dispersion and puncture-boundary spectral flow in the Hofstadter butterfly}
\label{subsec:spectral_flow}

The open-boundary spectra and wave-packet dynamics above identify the anomalous Floquet regime through states on the outer boundary of a finite hyperbolic patch. We now give a complementary diagnostic that avoids the extensive outer boundary altogether. This diagnostic isolates boundary behavior in hyperbolic systems where extensive boundaries and the lack of conventional momentum-space tools complicate Euclidean-style diagnostics.

We start from a compact periodic $\{8,3\}$ lattice with $N=2048$ sites, chosen from the list of trivalent symmetric graphs of Conder and Dobcs\'anyi~\cite{conder_trivalent_2002}, and introduce a small internal boundary by deleting a single vertex and its three incident edges. The unpunctured graph is a periodic $\{8,3\}$ lattice with no open boundary, while the punctured graph has $2047$ sites and a controlled local boundary.

\begin{figure*}[ht]
    \centering
    \includegraphics[height=1.55in, valign=t]{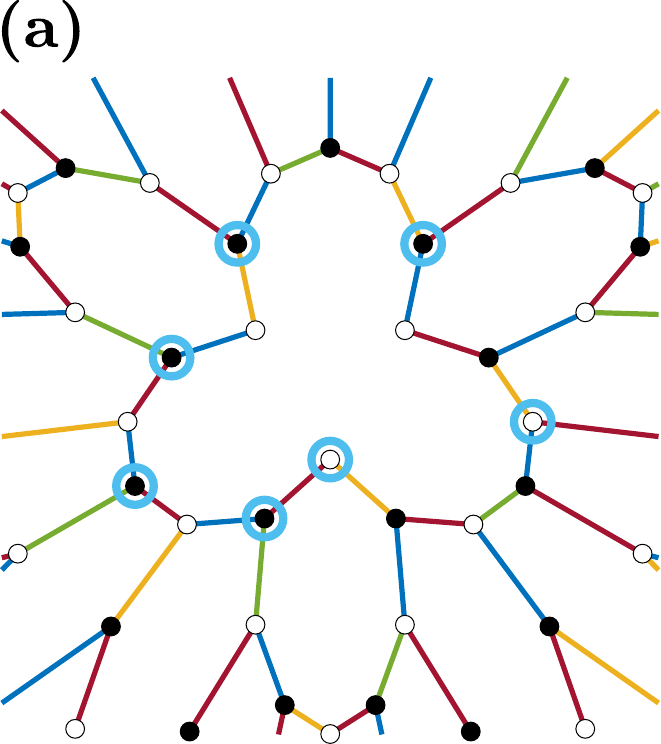}
    \hspace{0.1in}
    \includegraphics[height=1.89in, valign=t]{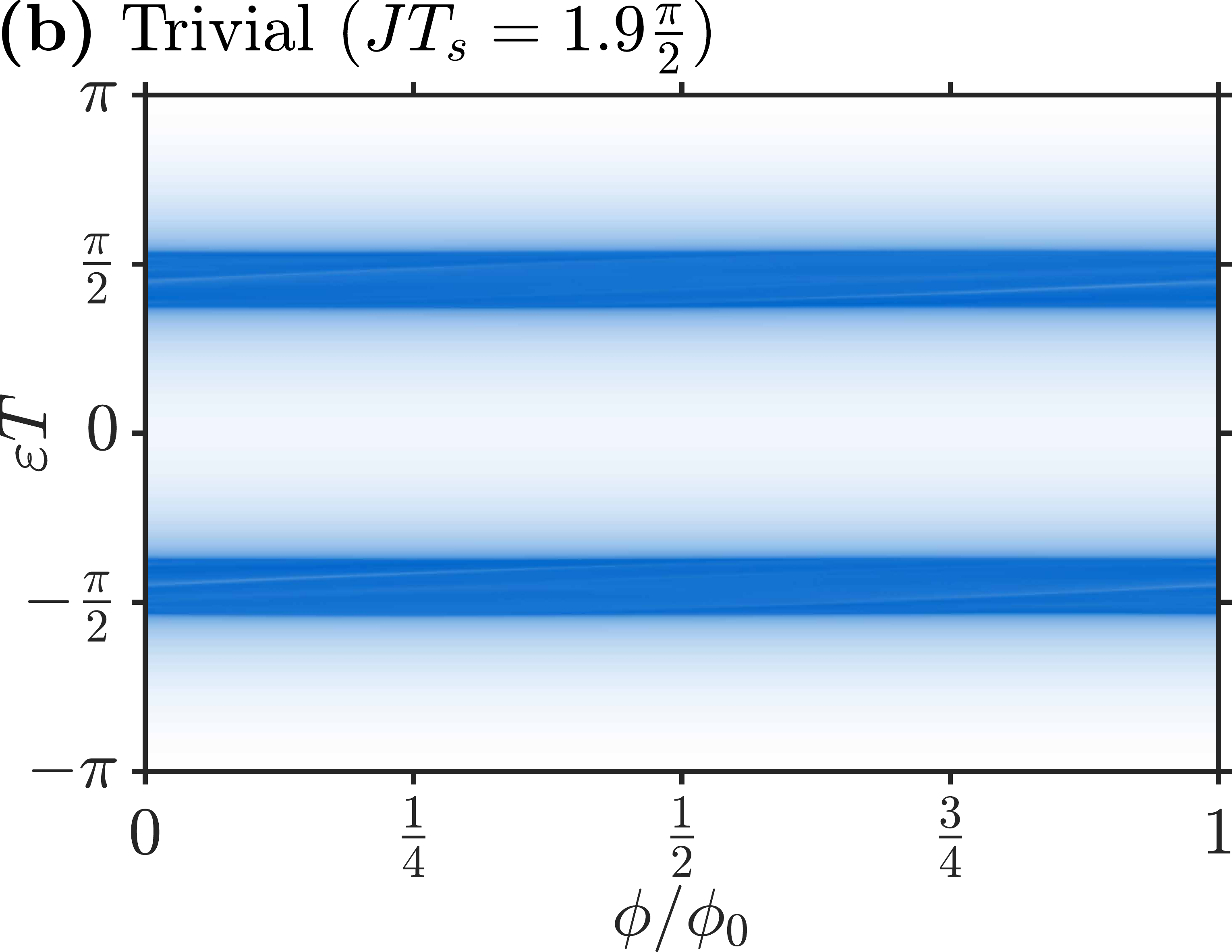}
    \hspace{0.1in}
    \includegraphics[height=1.89in, valign=t]{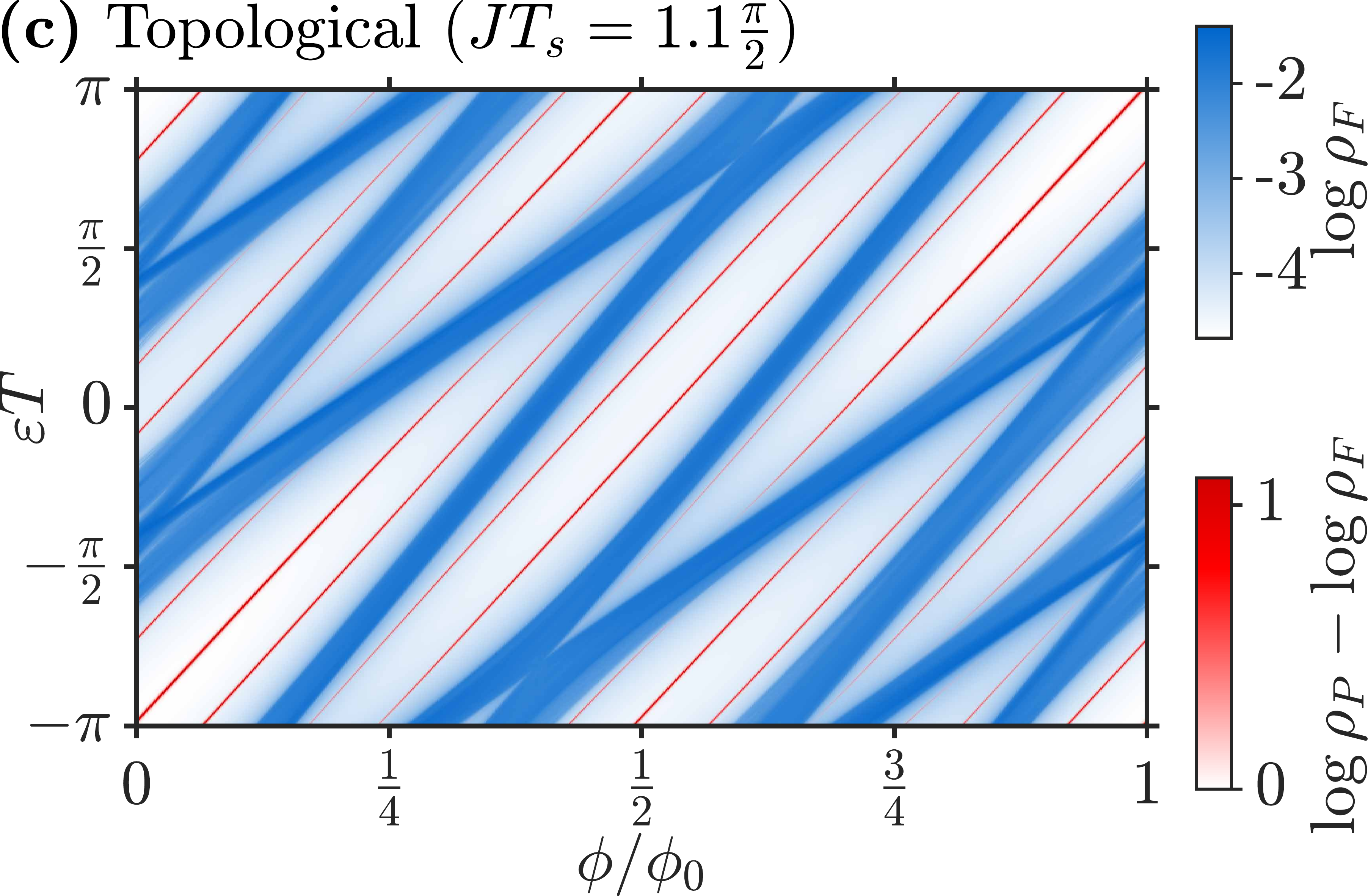}
    \caption{
    \textit{Flux dispersion and puncture-boundary spectral flow in compact periodic $\{8,3\}$ lattices.} (a) Local geometry of the puncture in a $2048$-site periodic lattice. A single vertex and its three incident edges are removed, creating a small boundary associated with the three octagons meeting at the deleted vertex. Cyan circles mark the seven puncture-boundary sites that cycle clockwise under one quarter of the hopping schedule at the perfect-hopping point. (b) Hofstadter density of states map in the trivial regime, $J\Tstep=1.9\pi/2$, with $\delta \Tstep=0.8 \pi/2$. The blue background is the logarithmic density of states of the full periodic lattice, while the red overlay shows the puncture-induced excess density of states. No in-gap puncture-boundary branches are visible. (c) Corresponding map in the anomalous Floquet regime, $J\Tstep=1.1\pi/2$, with $\delta \Tstep=0.8 \pi/2$. The puncture produces seven boundary branches that traverse the quasienergy gaps and hybridize with the bulk bands where they overlap. The colorbars on the right apply to both (b) and (c).
    }
    \label{fig:spectral_flow}
\end{figure*}

The local puncture geometry is shown in Fig.~\ref{fig:spectral_flow}(a). The removed vertex is shared by three octagons. At the perfect-hopping point, seven sites around the puncture, marked by cyan circles, form a closed puncture-boundary cycle. During one blue--green--red--orange quarter of the hopping schedule, each of these seven sites hops to the next cyan-circled site in the clockwise direction. Other nearby degree-deficient sites execute closed clockwise loops around intact octagons at the perfect point and do not contribute to this propagating puncture-boundary cycle. Thus the puncture creates a small, well-controlled version of the same boundary physics seen on the outer edge of the finite patches. In the fixed orientation of Fig.~\ref{fig:spectral_flow}(a), both the seven-site puncture-boundary cycle and the local octagon loops circulate clockwise. This contrasts with the counterclockwise motion at the outer boundary of an open patch and reflects the fact that the puncture forms an internal, rather than external, boundary.

We turn this puncture construction into a Hofstadter spectral-flow diagnostic by threading a uniform magnetic flux through the compact periodic lattice. Flux-dependent gap traversal in Hofstadter spectra provides a useful probe of global topology, including for Floquet quasienergy spectra~\cite{asboth_spectral_2017}. We use a uniform Hofstadter flux $\phi$ per octagonal plaquette, implemented through Peierls phases following the construction used for hyperbolic Hofstadter butterflies in Ref.~\cite{stegmaier_universality_2022}. Details of the phase assignment, including our choice of vanishing AB phases along a fixed basis of noncontractible cycles, are given in Appendix~\ref{app:peierls_phases}. The unpunctured $2048$-site lattice has 768 octagonal faces, so the allowed uniform fluxes are $\phi/\phi_0=n/768$, with $n=0,\ldots,768$, where $\phi_0$ is the magnetic flux quantum. For each flux value we compute the quasienergy density of states of the full periodic lattice, $\rho_F(\eps T,\phi)$, and of the punctured lattice, $\rho_P(\eps T,\phi)$, using the same Peierls gauge and normalization. These densities of states are normalized per site and obtained with a Lorentzian broadening with ${\eta=8\times10^{-3}}$. In Figs.~\ref{fig:spectral_flow}(b) and \ref{fig:spectral_flow}(c), the blue background shows $\log \rho_F$. The red overlay shows the positive part of the difference,
\begin{equation}
    \left[\log \rho_P(\eps T,\phi)
    -\log \rho_F(\eps T,\phi)\right]_+ ,
\end{equation}
and therefore isolates spectral weight corresponding to the boundary states, if any. For completeness, the corresponding $\log \rho_F$ and $\log \rho_P$ maps are shown separately in Appendix~\ref{app:separate_hofstadter_maps}.

Figures~\ref{fig:spectral_flow}(b) and \ref{fig:spectral_flow}(c) show the resulting Hofstadter butterflies for representative points in the trivial and topological regimes, respectively. For visual clarity, we use points with $\delta \Tstep=0.8\pi/2$ and ${J\Tstep=1.1}$ and $1.9\pi/2$ that lie in the same gapped regions as the blue square and circle in Fig.~\ref{fig:phase_diagram}, rather than the exact parameter points used for the open-boundary spectra in Sec.~\ref{subsec:dos}. At these nearby points, the spectral flow is more clearly resolved. In the trivial regime [Fig.~\ref{fig:spectral_flow}(b)] the bulk bands are nearly horizontal as the flux is varied, and the puncture does not introduce visible in-gap spectral weight. This is consistent with the open-boundary density of states in Fig.~\ref{fig:dos}, where the same parameter regime has empty quasienergy gaps. 

In the anomalous Floquet regime [Fig.~\ref{fig:spectral_flow}(c)] the blue bulk bands themselves disperse with flux. This flux dispersion reflects the local clockwise loops of the bulk states near the perfect-hopping point: as shown by the purple-circled bulk vertices in Fig.~\ref{fig:model}, two of the four inequivalent vertex types undergo two hops during each quarter of the hopping schedule and therefore complete one clockwise loop of an octagon over the full hopping schedule. The other two vertex types undergo four hops per quarter and complete two clockwise loops. Accordingly, the blue spectrum contains two families of flux-dispersing bulk bands, associated with states that complete either one or two octagon loops per period. Since the latter enclose twice as much flux during their evolution, their quasienergy slopes are approximately twice those of the one-loop bands. At the perfect-hopping point, these slopes are $d(\varepsilon T)/d(\phi/\phi_0)=2\pi$ and $4\pi$, respectively; away from that point, the same two characteristic slopes remain visible except near band crossings. %

The red features in Fig.~\ref{fig:spectral_flow}(c) are qualitatively different. They are induced by the puncture and form seven boundary branches at a generic vertical cut through the Hofstadter map, except where they hybridize with the bulk bands. These branches traverse both the $0$ and $\pi$ quasienergy gaps, providing a spectral-flow diagnostic of the puncture-boundary states. Their number and slope can be understood directly from the perfect-point motion in Fig.~\ref{fig:spectral_flow}(a). Label the seven puncture-boundary sites by $j=0,\ldots,6$. At the perfect-hopping point, one full Floquet period advances a state by four sites along this seven-site cycle, up to site-dependent phases:
\begin{equation}
    U_{\partial}(\phi)|j\rangle
    =
    e^{-i\alpha_j(\phi)}
|j+4 \;\mathrm{mod}\; 7\rangle.
\end{equation}
The phases $\alpha_j$ need not be uniform; they include local hopping phases and the sublattice-staggered onsite step. After seven periods and $28=7\times4$ advances along the cycle, however, the state has visited all seven sites and returned to its starting point. The site-dependent, flux-independent part of the accumulated phase is therefore a single constant, which we denote by $\gamma_0$. During the same seven periods, the state winds four times around the puncture. Since the puncture encloses three octagons, the flux-dependent phase corresponds to $12\phi$. Thus
\begin{equation}
    U_{\partial}(\phi)^7
    =
    \exp\left[
    -i\left(
    \gamma_0
    +2\pi\,12\,\frac{\phi}{\phi_0}
    \right)
    \right]
    I_{\partial} ,
\end{equation}
with the sign fixed by the clockwise orientation in Fig.~\ref{fig:spectral_flow}(a). Here $I_{\partial}$ denotes the identity operator on the seven-site puncture-boundary subspace. The seven puncture-boundary quasienergy branches satisfy
\begin{equation}
    \eps_m T
    =
    \frac{\gamma_0}{7}
    +2\pi\frac{12}{7}\frac{\phi}{\phi_0}
    +\frac{2\pi m}{7}
    \pmod{2\pi},
    \quad m=0,\ldots,6.
\end{equation}
Consequently, away from hybridization with the bulk bands, the boundary branches are separated by $2\pi/7$ and have slope
\begin{equation}
    \frac{d(\eps T)}{d(\phi/\phi_0)}
    =
    2\pi\frac{12}{7}.
\end{equation}
The onsite step shifts the common intercept through $\gamma_0$, but does not change the spacing or the flux slope.

This puncture diagnostic connects the local perfect-point picture to the Floquet topology. In the trivial regime, creating a small boundary does not add in-gap spectral flow. In the anomalous Floquet regime, the puncture produces chiral branches in both the $0$ and $\pi$ gaps. The appearance of boundary states in both gaps with the same chirality is consistent with a Rudner-type anomalous phase rather than a Chern-band phase: in a two-band-group Floquet system, unequal net chiral content in the two gaps would indicate a nonzero Chern number for the intervening band group. Within the finite periodic systems studied here, we find no numerical evidence for such a separate Chern-band regime. Additional Hofstadter maps at representative points across the phase diagram are shown in Appendix~\ref{app:hofstadter_collage}. These maps do not reveal a robust region with unequal gap flow characteristic of a Chern-insulating phase; in the small apparent gapped regions away from the two main regimes, one of the two quasienergy gaps remains too small to support an unambiguous identification of a separate phase.

\section{Discussion and outlook}
\label{sec:discussion}

We have constructed a Rudner-type anomalous Floquet phase on the hyperbolic $\{8,3\}$ lattice. The model is based on a periodic four-coloring of the edges and a 17-step drive consisting of 16 hopping steps followed by a sublattice-staggered onsite pulse. Near the perfect-hopping point, where a single active bond transfers amplitude completely during one hopping step, the bulk dynamics reduces to local loops around octagons, while interrupted loops at a boundary become chiral boundary motion in the opposite direction. Near the perfect-hopping point, the same phase is visible through three complementary diagnostics: in-gap states in the open-boundary density of states, chiral wave-packet propagation along the outer boundary, and puncture-boundary spectral flow in a Hofstadter calculation.

The hyperbolic setting changes the role of boundaries in an essential way. In Euclidean systems, the boundary contribution becomes negligible compared with the bulk in the thermodynamic limit. Finite hyperbolic patches do not have this property: the number of boundary sites remains a finite fraction of the total system size. This makes boundary physics unusually prominent, which is advantageous for realizing and observing topological boundary modes. At the same time, it complicates the usual separation between bulk and boundary spectra. The compact periodic lattices used in the phase diagram and spectral-flow calculations address this problem by removing the outer boundary altogether. Puncturing such a lattice then introduces a small, controlled internal boundary, allowing the same chiral edge physics to be detected without relying on a finite flake whose boundary contains a large fraction of all sites.

The observed signatures are consistent with an anomalous Floquet phase rather than a Chern-band phase. In the topological regime, boundary states appear in both the $0$ and $\pi$ quasienergy gaps with the same chirality. In a two-band-group Floquet problem, the difference between the net chiral contents of the two gaps determines the Chern number of the intervening bulk band group. The equal gap content observed here is therefore consistent with vanishing Floquet-band Chern numbers and topology carried by the full time evolution, as in the Rudner model~\cite{rudner_anomalous_2013}. Within the finite periodic systems and parameter ranges studied here, we find no numerical evidence for a separate regime with unequal gap flow characteristic of Chern bands. We do not assign a numerical winding invariant in this work; instead, the anomalous character is diagnosed through the combined boundary, bulk-gap, and puncture-flow signatures.

The present work focuses on a clean Hamiltonian construction and on a set of numerical diagnostics that are directly tied to the perfect-hopping picture. This scope keeps the connection between local Floquet loops, interrupted boundary motion, and puncture-boundary spectral flow transparent. It also points to several natural extensions. The bulk phase diagram obtained from finite periodic lattices should be viewed as a practical finite-size bulk diagnostic: regions where the gap measure becomes small identify the parameter ranges in which the $0$ or $\pi$ gap closes within the numerical resolution of the quotient. Comparing different periodic lattices, especially those designed to converge in the sense of hyperbolic periodic boundary conditions constructions or supercell sequences~\cite{lux_converging_2023,lenggenhager_nonabelian_2023}, or using continued-fraction methods to compute bulk densities of states directly in large hyperbolic systems~\cite{mosseri_density_2023}, would sharpen this picture. Similarly, disorder, imperfect edge coloring, boundary roughness, and pulse-shape errors can be incorporated into the same open-boundary and puncture-based diagnostics; disorder is particularly interesting because it may also provide a route to disorder-induced Floquet topological phases and anomalous Floquet-Anderson regimes~\cite{titum_disorderinduced_2015,titum_anomalous_2016}. Because boundary sites form a finite fraction of a hyperbolic patch, such perturbations are not merely edge corrections but part of the central physics of hyperbolic Floquet topological systems.

A complementary theoretical direction is to formulate real-space invariants for Floquet phases on hyperbolic lattices. Real-space Chern markers and Bott indices have already proved useful for static topological phases without ordinary translational symmetry~\cite{bianco_mapping_2011,hastings_topological_2011,loring_disordered_2011}. Other many-body real-space approaches may also be relevant, including polarization-operator formulas in the spirit of Resta~\cite{resta_quantummechanical_1998} and swap-operator formulas that extract Chern numbers from a single many-body wave function without introducing a family of twisted boundary conditions~\cite{dehghani_extraction_2021}. The latter feature is especially appealing for hyperbolic lattices, where ordinary momentum-space and twist-angle constructions are less natural. These tools could provide direct diagnostics of Chern-type content in Floquet bands or eigenstates. The anomalous Floquet phase studied here, however, requires an invariant sensitive to the full periodized time evolution, not only to individual Floquet bands. Translation-independent bulk-edge indices and recent local or response-based formulations for Floquet systems provide promising starting points~\cite{graf_bulk_2018,ghosh_local_2024,peraltagavensky_streda_2025}. Extending these approaches to hyperbolic lattices would provide a bulk diagnostic independent of open-boundary spectra and help clarify Floquet bulk-boundary correspondence in geometries with extensive boundaries.

The coloring construction itself also suggests a broader class of Floquet models on hyperbolic lattices. The essential ingredient is not specific to the $\{8,3\}$ tiling, but to the existence of a compatible face coloring. For a trivalent lattice whose faces admit a proper three-coloring, one can convert the face coloring into a three-coloring of edges by assigning each edge the color not used by the two faces adjacent to it. Two of these edge colors can then be kept as independent hopping steps, while the third edge color can be split into two alternating colors. This produces the same four-color structure used here: two color classes form perfect matchings, and the remaining two together form a third matching. The resulting blue--green--red--orange protocol therefore gives a direct route to Rudner-type Floquet models on other face-three-colorable hyperbolic tilings. Exploring how the anomalous phase depends on the choice of tiling would clarify which features are universal and which are specific to the $\{8,3\}$ construction studied here.

\begin{acknowledgments}
We thank Yang-Zhi Chou, Huy Nguyen, Alexandra Behne, Kishor Bharti, Sheryl Mathew, and Michael Gullans for useful discussions. H.D.~acknowledges support from NSF Grant No.~OMA-2120757 and the Simons Foundation. A.F.~and A.V.G.~were supported in part by the NSF QLCI (award No.~OMA-2120757), DoE ASCR Quantum Testbed Pathfinder program (award No.~DE-SC0024220), NSF STAQ program,  AFOSR MURI, ONR MURI, ARL (W911NF-24-2-0107), and NQVL:QSTD:Pilot:FTL. A.F.~and A.V.G.~also acknowledge support from the U.S.~Department of Energy, Office of Science, National Quantum Information Science Research Centers, Quantum Systems Accelerator (award No.~DE-SCL0000121) and from the U.S.~Department of Energy, Office of Science, Accelerated Research in Quantum Computing, Fundamental Algorithmic Research toward Quantum Utility (FAR-Qu).
A.K. was supported in part by the NSF  (QLCI award No.~OMA-2120757 and award No.~PHY2047732), AFOSR (award No.~FA9550-21-1-0129).

\end{acknowledgments}

\bibliography{refs}

\onecolumngrid

\appendix

\setcounter{figure}{0} 
\setcounter{equation}{0}
\setcounter{section}{0}
\setcounter{tocdepth}{2}

\renewcommand{\figurename}{Figure}
\renewcommand{\thefigure}{A\arabic{figure}} 
\renewcommand{\theHfigure}{A.\arabic{figure}}
\renewcommand{\theequation}{A\arabic{equation}}
\renewcommand{\theHequation}{A.\arabic{equation}}

\section{Individual quasienergy gaps}\label{app:individual_gaps}

\begin{figure}[ht]
    \centering
    \includegraphics[width=5.5in]{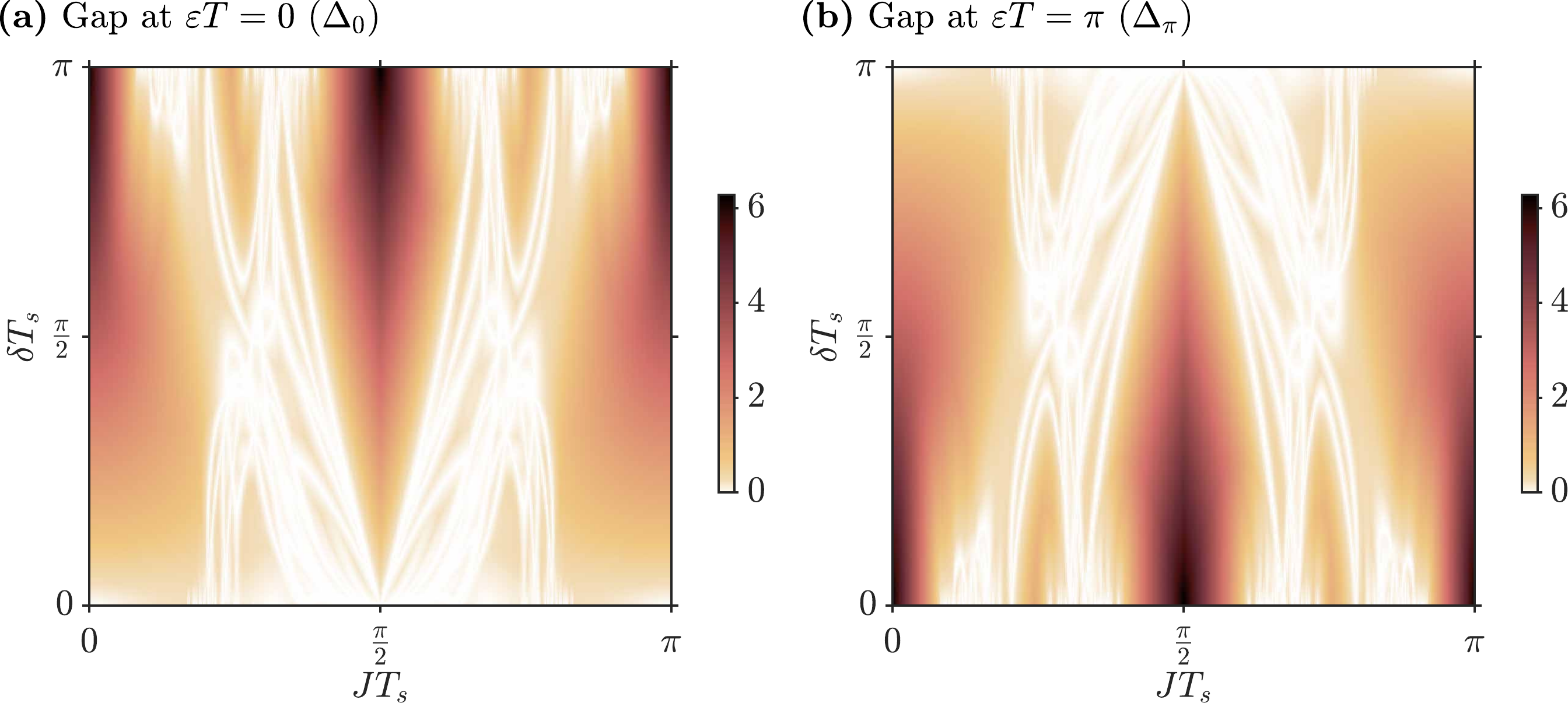}
    \caption{\textit{Individual bulk quasienergy gaps on the $2048$-site periodic $\{8,3\}$ lattice.} (a) $\Delta_0$ and (b) $\Delta_\pi$. Both gaps are shown in quasienergy-phase units over the same parameter range as Fig.~\ref{fig:phase_diagram}.}
    \label{fig:individual_gaps}
\end{figure}

In the main text, the bulk phase diagram is shown using the geometric mean $\bar{\Delta}=\sqrt{\Delta_0\Delta_\pi}$ of the quasienergy gaps at $\eps T=0$ and $\eps T=\pi$. This quantity is useful because it is nonzero only when both gaps are resolved in the finite periodic spectrum. For completeness, Fig.~\ref{fig:individual_gaps} shows the two gap measures separately, using the same 2048-site periodic $\{8,3\}$ lattice and the same parameter range as Fig.~\ref{fig:phase_diagram}. Small values of either $\Delta_0$ or $\Delta_\pi$ produce the low-gap regions in the geometric-mean phase diagram.

The gap maps in Fig.~\ref{fig:individual_gaps}, as well as the geometric-mean phase diagram in Fig.~\ref{fig:phase_diagram}, were obtained by diagonalizing the Floquet operator on a uniform $401\times401$ grid in the $(J\Tstep,\delta \Tstep)$ parameter plane.%

\section{Boundary- and bulk-state wave-packet dynamics}\label{app:wavepacket_dynamics}

\begin{figure}[h]
    \centering
    \includegraphics[width=2.1in]{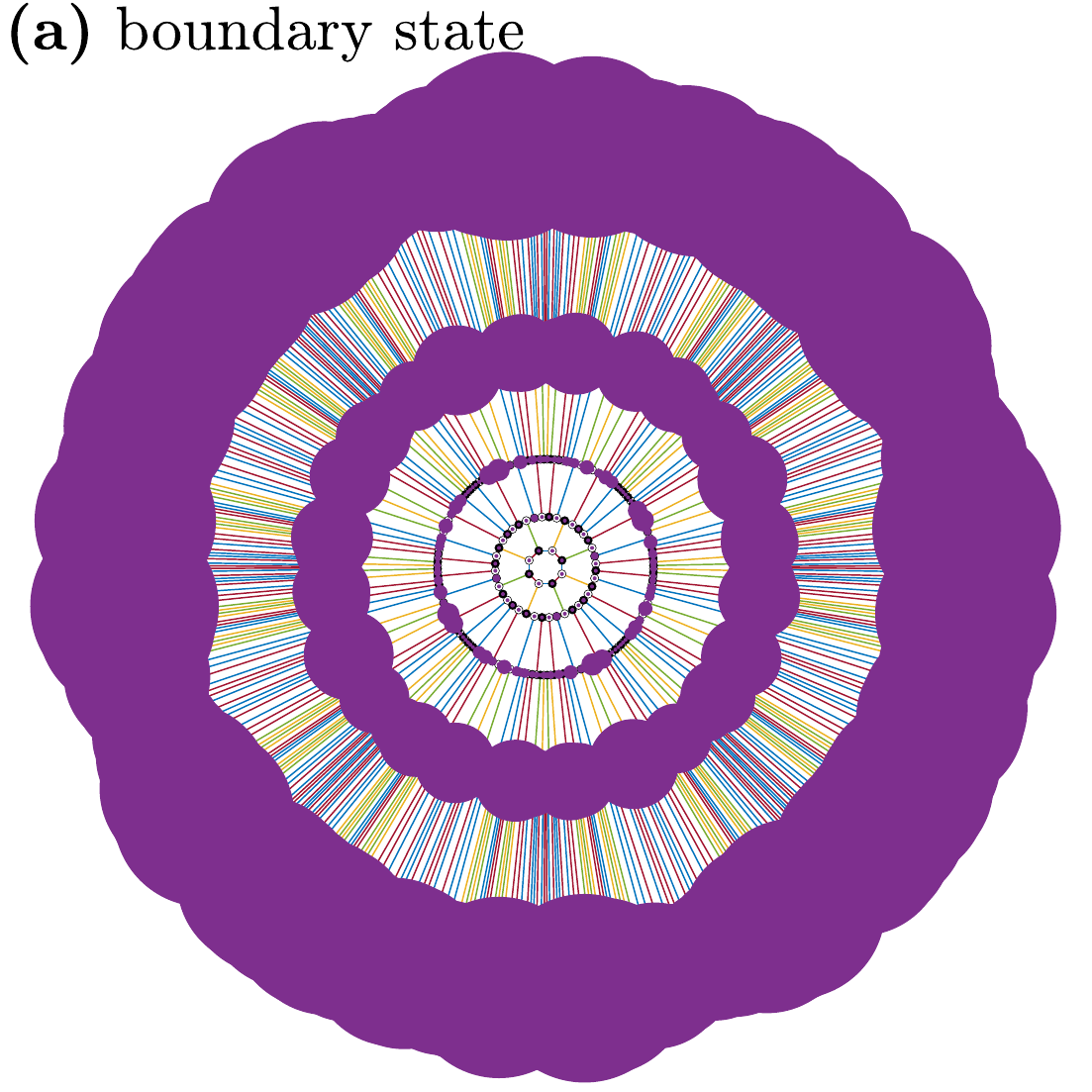}
    \includegraphics[width=2.1in]{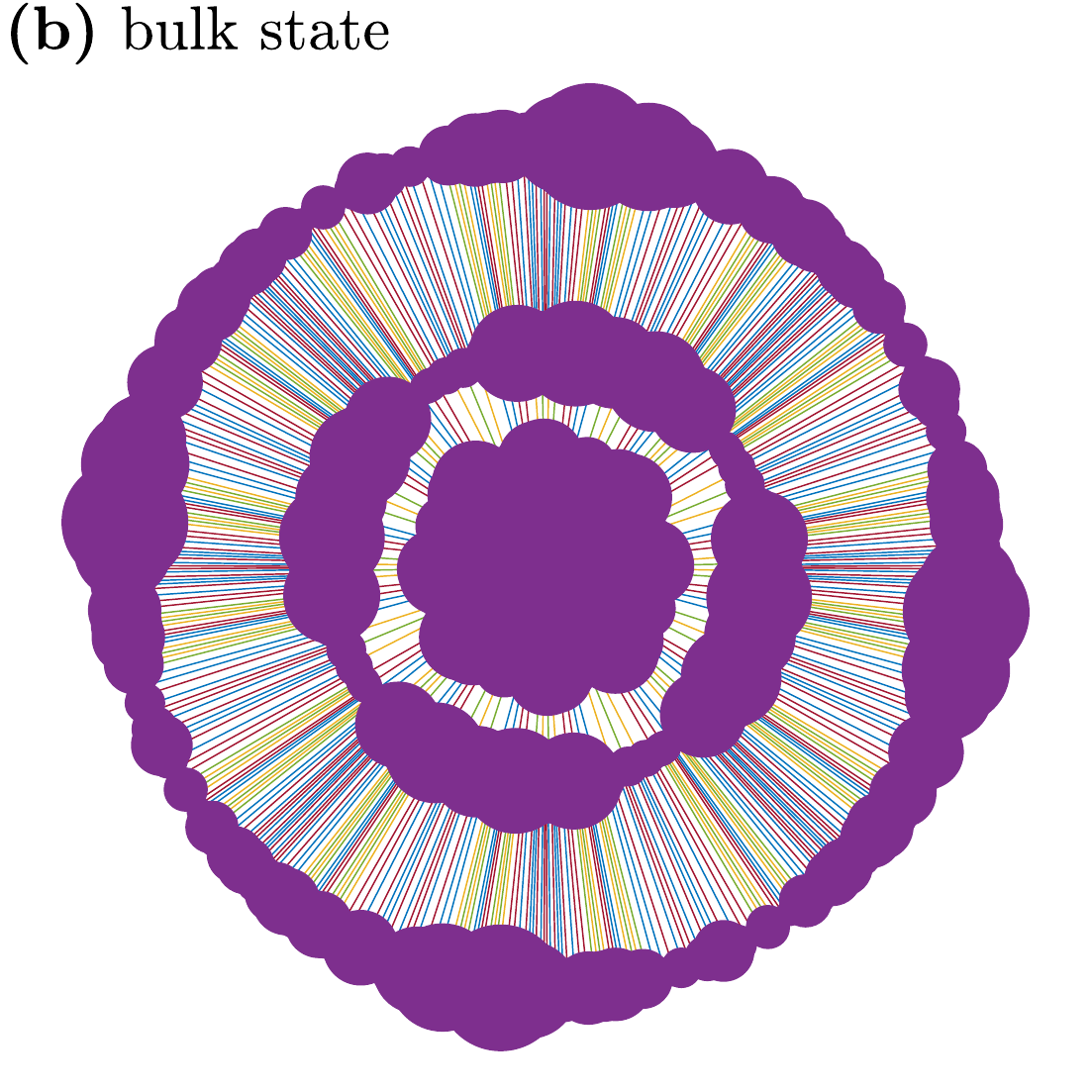}
    \caption{
    \textit{Floquet eigenstates used to construct the boundary and bulk wave packets.} Both states are shown on the five-shell open-boundary $\{8,3\}$ patch with $N=2888$ sites at $J\Tstep=1.1\pi/2$ and $\delta T_{1h}=\pi/4$. The sites belonging to each shell are plotted on a circle, as in Fig.~\ref{fig:boundary_dynamics}, and the area of each purple disk is proportional to the eigenstate probability $|\phi_i|^2$. (a) The in-gap boundary state whose quasienergy is marked in Fig.~\ref{fig:dos}. (b) A representative state from a bulk quasienergy band at $\eps T=1.3$. The states are shown before application of the angular Gaussian envelope in Eq.~\eqref{eq:initial_packet}. The disk areas are rescaled independently in each panel so that the largest disk has the same fixed display area. The disk sizes therefore show the relative spatial distribution within each state, but do not provide a direct comparison of absolute site probabilities between the two states.
    }
    \label{fig:boundary_bulk_states}
\end{figure}

In Sec.~\ref{subsec:boundary_dynamics}, the initial boundary wave packet is constructed by applying the angular Gaussian envelope in Eq.~\eqref{eq:initial_packet} to a Floquet eigenstate. Here we show the unfiltered eigenstate used in that construction and compare it with a representative state selected from a bulk quasienergy band. We also show the stroboscopic evolution of the wave packet constructed from the bulk-band state.

Figure~\ref{fig:boundary_bulk_states} shows the two normalized Floquet eigenstates before the angular envelope is applied. Both states are obtained from the same five-shell open-boundary $\{8,3\}$ patch with $N=2888$ sites and at the same anomalous-regime parameter point used for the boundary dynamics in the main text, $J\Tstep=1.1\pi/2$ and $\delta T_{1h}=\pi/4$. The boundary state in Fig.~\ref{fig:boundary_bulk_states}(a) is the in-gap state whose quasienergy is marked in the density of states in Fig.~\ref{fig:dos}. Its probability density is concentrated in the outer boundary region. The state in Fig.~\ref{fig:boundary_bulk_states}(b) is selected from a bulk quasienergy band at $\eps T=1.3$ and has appreciable weight over several shells. Here, ``bulk state'' refers to the spectral location of the state within a bulk band and does not imply localization near the geometric center of the finite hyperbolic patch.

For each seed eigenstate, we construct an initial wave packet using the same angular envelope as in Eq.~\eqref{eq:initial_packet} and normalize the resulting state. The labels ``boundary'' and ``bulk'' refer to the corresponding seed eigenstates; after multiplication by the spatial envelope, the resulting wave packets are no longer Floquet eigenstates. Because the envelope depends only on the angular coordinate, it localizes the state azimuthally while retaining the shell-resolved probability profile inherited from the seed eigenstate. The bulk wave packet therefore has a broad radial profile already at $t=0$.

\begin{figure}[h]
    \centering
    \includegraphics[width=0.95\textwidth]{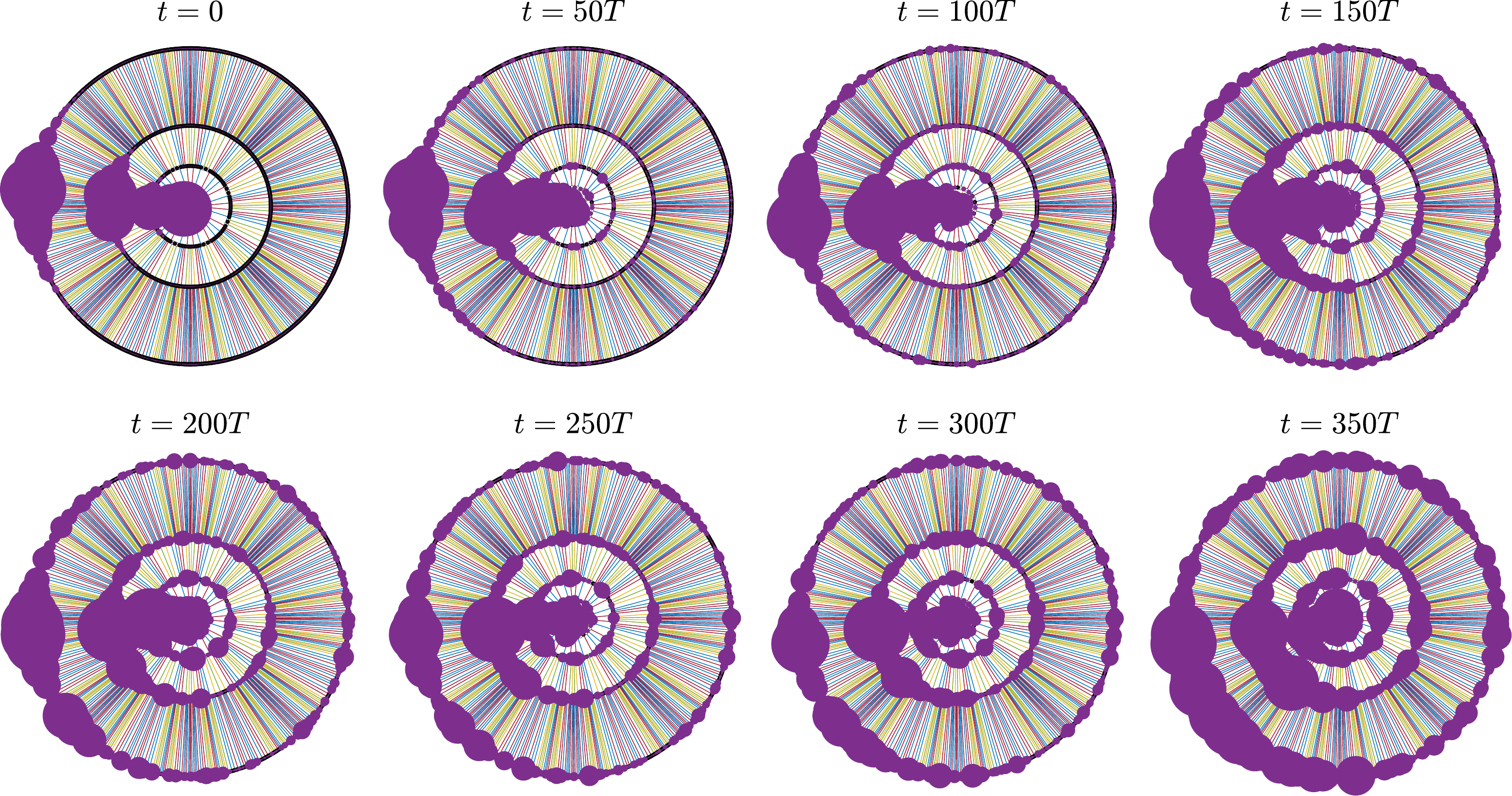}
    \caption{
    \textit{Stroboscopic evolution of a wave packet constructed from a bulk-band state in the anomalous Floquet regime.} The seed Floquet eigenstate is the state at $\eps T=1.3$ shown in Fig.~\ref{fig:boundary_bulk_states}(b). The system and drive parameters are the same as in Fig.~\ref{fig:boundary_dynamics}: a five-shell open-boundary $\{8,3\}$ patch with $N=2888$ sites, $J\Tstep=1.1\pi/2$, and $\delta T_{1h}=\pi/4$. The initial state is obtained by applying the angular Gaussian envelope in Eq.~\eqref{eq:initial_packet} and then normalizing the result. The area of purple disks are proportional to $|\psi_i(t)|^2$ and are rescaled independently at each time so that the largest marker has a fixed display area; disk sizes should therefore not be compared quantitatively across panels. The panels show $t=0,50T,100T,150T,200T,250T,300T,$ and $350T$. The packet develops probability over multiple shells and does not exhibit the persistent chiral boundary propagation seen in Fig.~\ref{fig:boundary_dynamics}.
    }
    \label{fig:bulk_dynamics}
\end{figure}

Figure~\ref{fig:bulk_dynamics} shows the subsequent stroboscopic evolution of the bulk wave packet. In contrast to the boundary wave packet in Fig.~\ref{fig:boundary_dynamics}, the bulk packet does not remain confined to the outer boundary region or translate around the sample as a compact packet with a definite chirality. Instead, it loses its initial angular localization and develops substantial probability around multiple shells. This comparison shows that the persistent counterclockwise propagation in Fig.~\ref{fig:boundary_dynamics} is a property of the in-gap boundary-state sector rather than a generic consequence of the angular localization procedure.

\section{Peierls phases in the Hofstadter calculations} \label{app:peierls_phases}

We implement a uniform magnetic flux on the compact periodic $\{8,3\}$ lattices using the Peierls substitution, following the construction of Ref.~\cite{stegmaier_universality_2022}. For an oriented edge $(i,j)$, we introduce a hopping phase $\varphi_{ij}=-\varphi_{ji}$ and replace the hopping Hamiltonian in each color sector by
\begin{equation}
H_\mu(\phi)
=
-J\sum_{\langle ij\rangle\in E_\mu}
\left(
e^{i\varphi_{ij}}c_i^\dagger c_j
+
e^{-i\varphi_{ij}}c_j^\dagger c_i
\right),
\qquad
\mu=b,g,r,o.
\end{equation}
The same hopping phase is used each time a given edge is activated during the Floquet drive, while the sublattice-staggered onsite step remains unchanged.

The hopping phases are chosen such that the oriented phase accumulated around every octagon $f$ is
\begin{equation}
    \sum_{\langle ij\rangle\in\partial f}\varphi_{ij}
    =
    2\pi\frac{\phi}{\phi_0}
    \pmod{2\pi},
\end{equation}
where the sum follows the edges clockwise around the octagon, $\phi$ is the magnetic flux through each octagon, and $\phi_0$ is the magnetic flux quantum. On a closed regular map, only $F-1$ of the $F$ octagon constraints are independent, and consistency requires
\begin{equation}
    F\frac{\phi}{\phi_0}\in\mathbb{Z}.
\end{equation}
The $2048$-site periodic lattice used in the Hofstadter calculations contains $F=768$ octagons. The inequivalent uniform flux values are therefore
\begin{equation}
    \frac{\phi}{\phi_0}=\frac{n}{768},
    \qquad n=0,\ldots,767.
\end{equation}
We additionally include $n=768$ in the plots to display the endpoint $\phi/\phi_0=1$, which is equivalent to zero flux.

The octagon constraints do not completely determine the hopping phases on a compact surface. In addition to the $F-1$ independent octagon cycles, a genus-$g$ regular map has $2g$ independent noncontractible Aharonov--Bohm (AB) cycles. Together, these cycles form a basis of the cycle space. We choose representative noncontractible cycles $\mathcal{C}_a$ and specify their AB phases through
\begin{equation}
    \sum_{\langle ij\rangle\in\mathcal{C}_a}\varphi_{ij}
    =
    k_a
    \pmod{2\pi},
    \qquad a=1,\ldots,2g.
\end{equation}
A local gauge can then be fixed, for example, by setting the hopping phases on the edges of a spanning tree.

For the calculations presented here, we use a single AB-flux configuration and set
\begin{equation}
    k_a=0,
    \qquad a=1,\ldots,2g,
\end{equation}
for the chosen representatives of the noncontractible cycles. Once the octagon fluxes and AB holonomies are fixed, a local gauge transformation may redistribute the hopping phases among individual edges but does not change the quasienergy spectrum or any gauge-invariant observable. The Floquet eigenstates in different gauges are related by site-dependent phase rotations. At nonzero octagon flux, a deformation of a noncontractible-cycle representative by adding octagon boundaries also adds the flux enclosed by those octagons. Consequently, changing the cycle representatives requires a corresponding redefinition of the AB phases in order to preserve the same physical holonomies. Our choice $k_a=0$ should therefore be understood relative to the fixed cycle basis used in the numerical calculation.

Unlike Ref.~\cite{stegmaier_universality_2022}, where multiple AB-flux configurations are sampled to obtain a denser approximation to the infinite-lattice spectrum, we retain this single zero-AB-phase configuration throughout. The same hopping-phase assignment is used for the full and punctured periodic lattices.

For the punctured calculation, the Peierls phases are first assigned on the full regular map. We then remove one vertex and its three incident edges, leaving the phases on all remaining edges unchanged. No additional AB flux is inserted through the puncture. The puncture-boundary cycle encloses the three octagons adjacent to the removed vertex, so its accumulated magnetic phase is
\begin{equation}
    \varphi_{\mathrm{puncture}}
    =
    2\pi\frac{3\phi}{\phi_0}
    \pmod{2\pi}.
\end{equation}
This gives the factor of three in the puncture-boundary spectral-flow slope discussed in Sec.~\ref{subsec:spectral_flow}.

\section{Separate full and punctured Hofstadter density of states maps}\label{app:separate_hofstadter_maps}

\begin{figure}[ht]
    \centering
    \includegraphics[width=0.8\textwidth]{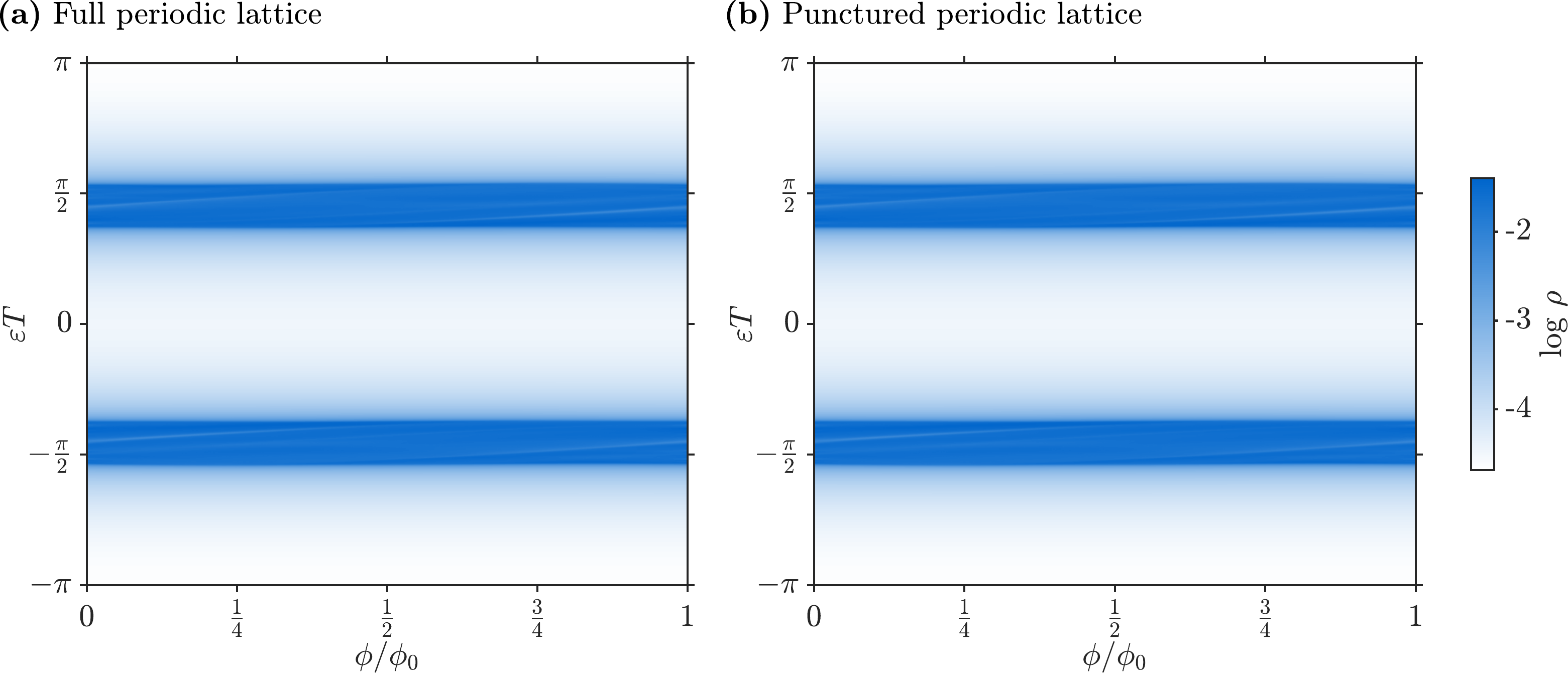}
    \caption{\textit{Separate Hofstadter density of states maps in the trivial regime, $J\Tstep=1.9\pi/2$ and $\delta \Tstep=0.8\pi/2$.} (a) Full 2048-site periodic lattice.  (b) Punctured periodic lattice obtained by removing one vertex and its three incident edges.}
    \label{fig:trivial_full_punctured_dos}
\end{figure}

\begin{figure}[ht]
    \centering
    \includegraphics[width=0.8\textwidth]{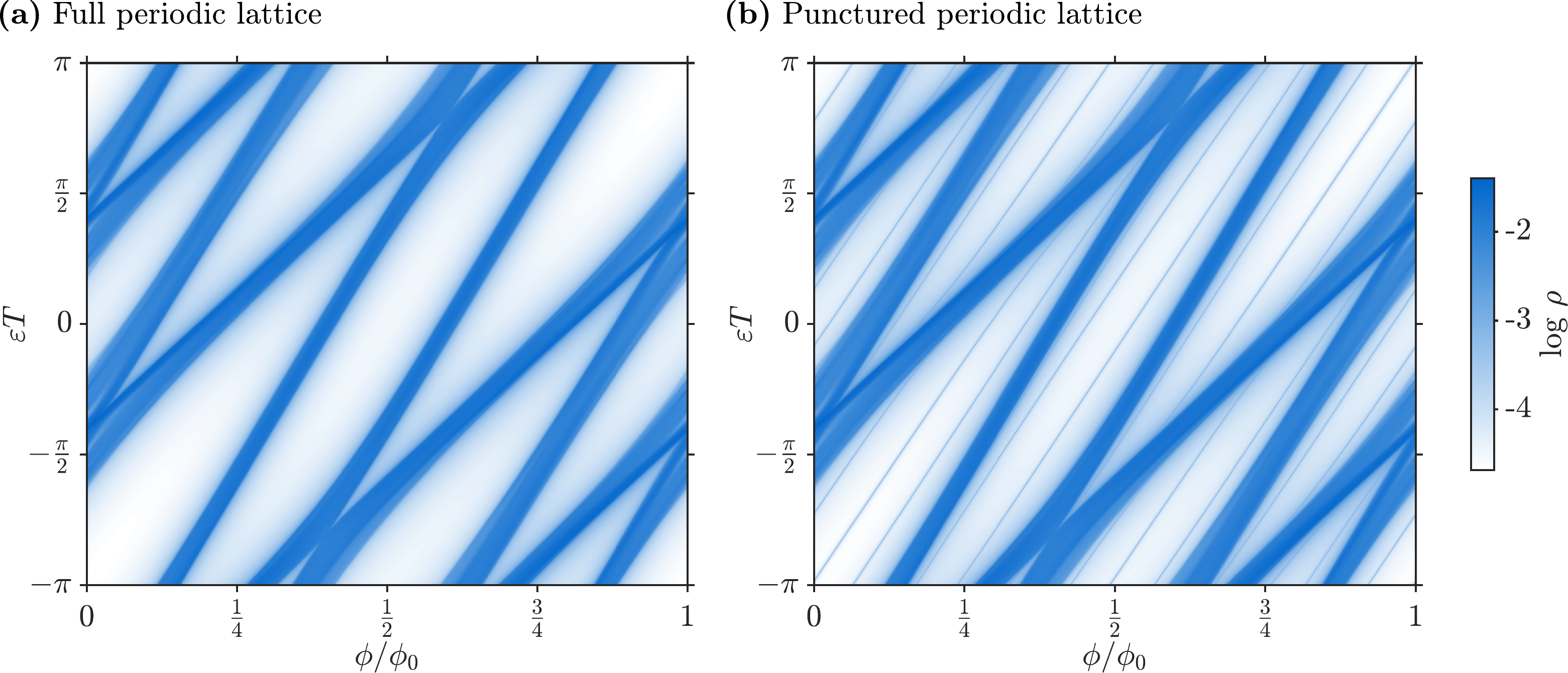}
    \caption{\textit{Separate Hofstadter density of states maps in the anomalous Floquet regime, $J\Tstep=1.1\pi/2$ and $\delta \Tstep=0.8\pi/2$.} (a) Full 2048-site periodic lattice. (b) Punctured periodic lattice obtained by removing one vertex and its three incident edges.}
    \label{fig:topological_full_punctured_dos}
\end{figure}

In the main text, the Hofstadter maps are shown as overlays in Fig.~\ref{fig:spectral_flow}: the full periodic-lattice density of states is plotted in blue, while the puncture-induced excess density of states is plotted in red. For completeness, Figs.~\ref{fig:trivial_full_punctured_dos} and \ref{fig:topological_full_punctured_dos} show the full and punctured density of states maps separately. The parameters are the same as in Fig.~\ref{fig:spectral_flow}. Comparing the two panels in each figure shows directly which spectral features are already present in the full periodic lattice and which ones are introduced by the puncture. While the bulk density of states is essentially unchanged by the puncture in both regimes, only the anomalous Floquet regime develops puncture-induced boundary states that traverse the bulk quasienergy gaps.

\section{Additional Hofstadter maps across the phase diagram}\label{app:hofstadter_collage}

\begin{figure}[ht]
    \centering
    \includegraphics[width=\textwidth]{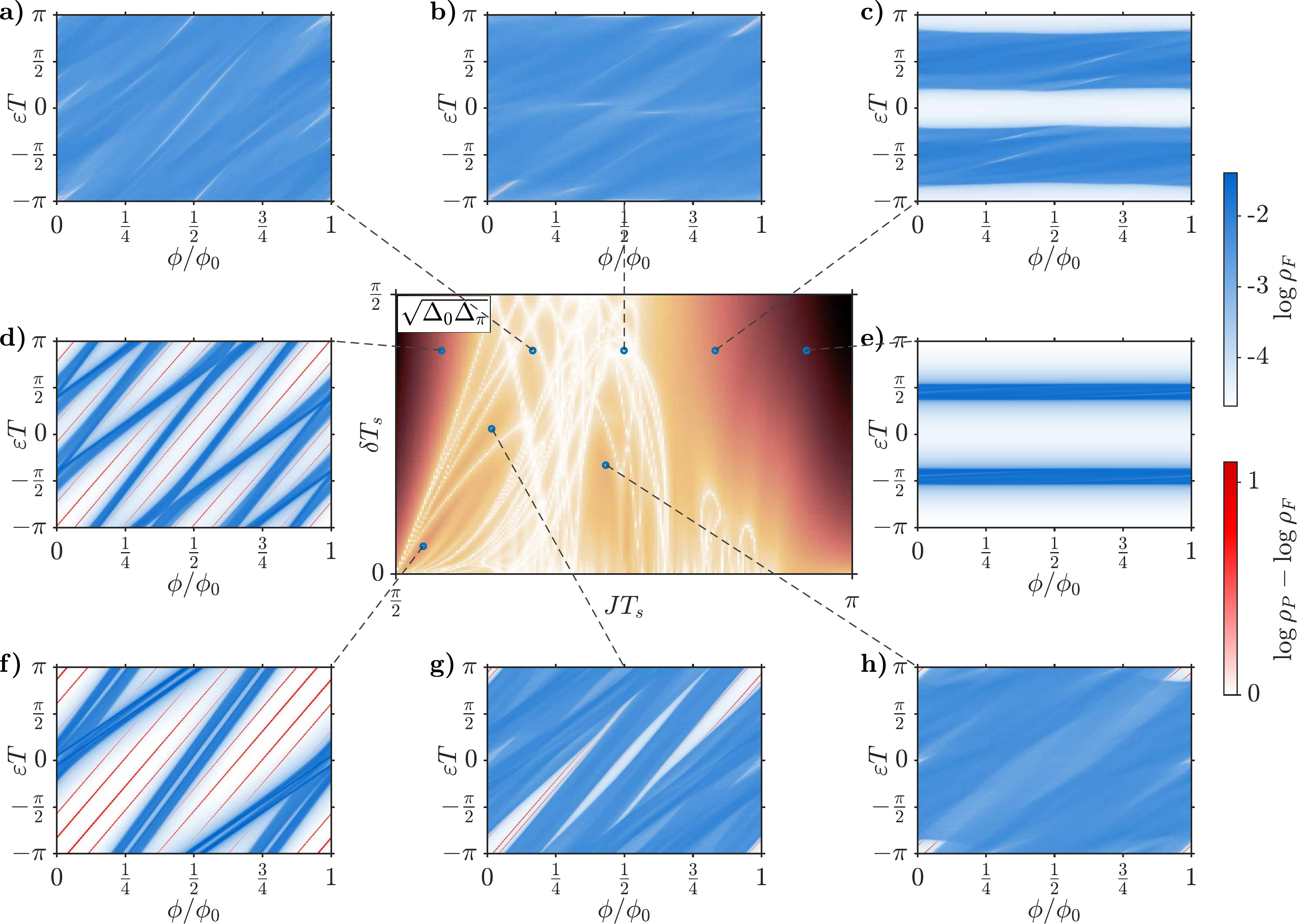}
    \caption{\textit{Additional Hofstadter density of states maps across the phase diagram.} The central panel shows a zoomed-in geometric-mean gap diagram, $\bar{\Delta}=\sqrt{\Delta_0\Delta_\pi}$, with markers indicating the parameters of panels (a)--(h). The surrounding panels show Hofstadter maps with the same full-lattice density of states and puncture-induced excess-density convention used in the main text. The blue background is the logarithmic density of states of the full periodic lattice, while the red overlay shows the puncture-induced excess density of states. The colorbars on the right apply to all Hofstadter maps.}
    \label{fig:hofstadter_collage}
\end{figure}

In the main text, the Hofstadter density of states maps are shown for two representative points: the anomalous Floquet regime and the trivial regime. For completeness, Fig.~\ref{fig:hofstadter_collage} shows additional Hofstadter maps at several points in the bulk gap phase diagram. The central panel is a zoomed-in version of the geometric-mean gap diagram $\bar{\Delta}=\sqrt{\Delta_0\Delta_\pi}$ shown in Fig.~\ref{fig:phase_diagram}, with markers indicating the parameters used for the surrounding density of states maps. The central panel covers the region $J\Tstep\in[\pi/2,\pi]$ and $\delta \Tstep\in[0,\pi/2]$. The remainder of the phase diagram in Fig.~\ref{fig:phase_diagram} is obtained by mirror-symmetric extension of the displayed region.

Panels~\ref{fig:hofstadter_collage}(d) and \ref{fig:hofstadter_collage}(e) correspond to the two representative points discussed in the main text, with $J\Tstep=1.1\pi/2$ and $J\Tstep=1.9\pi/2$, respectively, at fixed $\delta \Tstep=0.8\pi/2$. Panels~\ref{fig:hofstadter_collage}(a)--(c) show intermediate values, $J\Tstep=1.3\pi/2$, $1.5\pi/2$, and $1.7\pi/2$, at the same value of $\delta \Tstep$, illustrating how the Hofstadter spectrum evolves between the anomalous and trivial regimes.

Panels~\ref{fig:hofstadter_collage}(f)--(h) show three additional points where the finite-size gap diagnostic indicates an apparent isolated region of nonzero $\bar{\Delta}$. These points are $(J\Tstep,\delta \Tstep)=(1.06\pi/2,0.10\pi/2)$, $(1.21\pi/2,0.52\pi/2)$, and $(1.46\pi/2,0.39\pi/2)$, respectively. In these cases, one of the two quasienergy gaps is too small to support a clear identification of a separate phase, while the other gap shows puncture-induced boundary states. We therefore do not assign these regions to distinct Chern-insulating phases.

\end{document}